\documentclass[lettersize,journal]{IEEEtran}
\usepackage{cite}
\usepackage{graphicx}
\usepackage{amsmath}
\usepackage{amssymb}
\usepackage{bm}
\usepackage{url}
\usepackage{thm}
\newtheorem{theorem}{Theorem}
\newtheorem{lemma}{Lemma}

\begin{document}
\title{Estimating the number of superimposed sinusoids}

\author{Aleksandr~Kharin,~\IEEEmembership{Member,~IEEE}
\thanks{
{A. Kharin was with the Laboratoire de Math\'ematiques Rapha\"el Salem, Universit\'e de Rouen-Normandie, Rouen, France (e-mail: aleksandr.kharin.1989@gmail.com, kharin@ieee.org ORCID: orcid.org/0000-0002-4534-7046).}}
\thanks{Manuscript received XX XX, 2020. This work was supported by the French RIN project FUMA}}

\markboth{Journal of \LaTeX\ Class Files,~Vol.~14, No.~8, August~2021}%
{Shell \MakeLowercase{\textit{et al.}}: A Sample Article Using IEEEtran.cls for IEEE Journals}

\maketitle

\begin{abstract}
Estimation of the number of superimposed sinusoids in the presence of noise is an important model order selection (MOS) problem in statistical signal processing.  In this paper, we propose a new approach to the design of MOS algorithms for estimating the number of superimposed sinusoids. Our  proposed approach is partially based on the minimum error probability criterion.
Also, we pay a lot of attention to the performance and consistency analysis of the MOS algorithms. In this study, an error probability is used as a universal performance measure of the MOS algorithms. We propose a theoretical framework that makes it possible to provide consistency analysis and to obtain closed-form expressions for the approximated error probabilities of a wide range of MOS algorithms. As an example, we applied this framework to the consistency and performance analysis of several MOS algorithms for estimating the number of superimposed sinusoids. Using the obtained results, we provide a parametric optimization of the presented MOS algorithms. 
Finally, we examine a quasilikelihood approach to the design and performance analysis of the MOS algorithms. The proposed theoretical framework is used to find the scope of the quasilikelihood approach.

\end{abstract}

\begin{IEEEkeywords}
model order selection, sinusoids in noise, superimposed sinusoids, minimum error probability criterion, abridged error probability, number of signals estimating, maximum likelihood, penalty term, quasilikelihood approach.
\end{IEEEkeywords}

\IEEEpeerreviewmaketitle

\section{Introduction} \label{myintro}
\IEEEPARstart{I}{n} this paper, we study the problem of estimating the number of superimposed sinusoids. This important kind of model order selection (MOS) problem \cite{Stoica2004, SpectralApp1, MS1} arises in signal processing and its applications such as radar and sonar theory \cite{RadarApp1, RadarApp2}, communication theory \cite{CommApp1}, astronomy \cite{RAapp}, spectral analysis \cite{SpectralApp1}, and many others.

Modern MOS algorithms for estimating the number of superimposed sinusoids are based on various optimality criteria. For example, the Akaike information criterion (AIC) \cite{AIC}, the Bayesian information criterion (BIC) \cite{BIC, BIC2}, the Minimum Description Length (MDL) criterion \cite{MDL}, the maximum a posteriori criterion (MAP) \cite{MAP}, the Exponentially embedded families (EEF) criterion \cite{EEF, Kay2, EEF2018}, the Penalizing adaptively likelihood (PAL) criterion \cite{PAL}, Residual Ratio Thresholding (RRT) \cite{RRT} etc. Good surveys on the MOS criteria can be found, for instance, in \cite{Stoica2004}, \cite{7465796}, and \cite{MS1}. Also, there are some lesser-known approaches, see, for example, the nonparametric MOS procedure in \cite{nonparam} and the tensor approach in \cite{tensor1, tensor2}. In some rare cases, the maximum likelihood method can be directly used to estimate the model order, for example, \cite{TrifonovKharin2, TrifonovKharin3, TrifonovKharin5, TrifonovKharin6}. The MOS criteria define the structure of the corresponding algorithms, specifically, these criteria define the {\it penalty terms} of the corresponding algorithms.

The problem of estimating the number of superimposed sinusoids is a {\it parametric} MOS problem \cite{MS1}. In a typical parametric MOS problem, the {\it true} model is the most appropriate model \cite{MS1}. Hence, in this paper, we formulate the problem of estimating the number of superimposed sinusoids as a problem of finding the {\it true} number of superimposed sinusoids.  In this case, the probability of the event that the estimation of the model order is not equal to the true model order (error probability) can be used as a performance measure of the MOS algorithms. One can equivalently use the probability of the complementary event known as the probability of correct detection (or correct estimation or correct model order estimation, etc). The error probability and other equivalent performance measures are very important for the problem of estimating the number of superimposed sinusoids  \cite{MAP, EEF, Kay2, Djuric1996, 7465796, tensor1, tensor2, Mariani2015, Nadler2011}.

If the goal is to find the true model order (true number of superimposed sinusoids), then it is also important to minimize the error probability. However, most existing MOS criteria do not minimize the error probability. In this paper, we propose a new approach to designing MOS algorithms. This approach is partially based on the minimum error probability criterion. Hence, MOS algorithms based on the proposed approach have better performance (in terms of the error probability) compared to algorithms based on the previously known criteria.

A significant part of this work is devoted to the performance analysis and consistency analysis of the MOS algorithms. We use the conception of the error probability to provide a universal performance measure for MOS algorithms. Next, we develop theoretical and numerical frameworks that make it possible to calculate this performance measure for a wide range of MOS algorithms. Also, the provided results allow to obtain consistency conditions for a wide range of MOS algorithms. 

The problem of estimating the number of superimposed sinusoids involves the use of maximum likelihood (ML) frequency estimates in the MOS algorithms. This fact significantly increases the computational complexity of the MOS algorithms. Thus, we propose the so-called quasilikelihood (QL) approach to the design of the MOS algorithms. On the one hand, the proposed QL design approach can solve the mentioned computational complexity problem, but on the other hand, this approach has some limitations. We use the developed theoretical and numerical frameworks to find these limitations and justify the use of QL design approach. Moreover, the proposed theoretical framework makes it possible to determine whether a MOS algorithm is robust to errors in the used frequency values.

The paper is organized as follows. Section \ref{ProbForm} presents the formulation of the problem. Section \ref{algdes} covers the modern penalty terms considered in this paper and presents a new approach to the design of MOS algorithms. A theoretical framework for the consistency and performance analysis is described in Section \ref{perfan}. In Section \ref{BLrob}, we collect the results connected with the QL approach. We present the main problem of one type of MOS algorithms based on the proposed design approach in Section \ref{PMEPfuture}, also, we present some possible solutions to this problem. Moreover, in Section VI, we bring together all the results previously obtained in this paper and formulate ways of future developing the proposed approach to the design of MOS algorithms. Section \ref{sims} presents numerical results supplementing the theoretical ones.
\section{Problem Formulation} \label{ProbForm}
We start from the general formulation of the problem of estimating the number of superimposed signals.
Let the observed data $x(t)$ be a signal composed of an unknown number of signals ${\rm \nu } $ at unknown parameters, and corrupted by the additive white Gaussian noise (AWGN). Thus, one can represent $x(t)$ as
\begin{equation} \label{SModel} 
x(t) = s_{\nu_0}(t,\bm{\Theta}_{\nu_0}) + \sigma_0 n(t),\,\,\,\, s_\nu(t,\bm{\Theta}_\nu) =\sum\limits_{i = 1}^{\rm{\nu }} {{s_i}(t,{{\bm{\theta}}_{i}})},
\end{equation} 
where $t \in \left\lbrace t_1, \ldots t_{N_s} \right\rbrace$, $N_s$ denotes the number of samples; ${\nu } \in \{1,..,\mathcal{N}\}$ denotes the number of signals; ${{\bm{\theta}}_i} =\left(\theta_{i1} ,...,\theta_{i{\rm \mu }_{i} } \right)$ is a vector of the unknown parameters of the $i$-th signal,  ${\rm \mu }_{{ i}} $ is the number of the unknown parameters of the $i$-th signal, $\bm{\Theta}_\nu = \left({{\bm{\theta}}_1} ,...,{{\bm{\theta}}_\nu}  \right)$; $n(t)$ is the AWGN, i.e., $\left\lbrace  n(t) \right\rbrace_{t=1}^{N_s}$ are independent identically distributed Gaussian random variables with zero means and unit variances; $\sigma$ denotes a noise level (noise variance). We denote the true values of the unknown parameters by subscript $0$. 

The problem is to estimate the unknown number of signals using the observed data $x(t)$. This problem can also be interpreted as a MOS problem.

An algorithm of estimating the number of signals (hereafter also called a MOS algorithm) is essentially a function from the space of observed signals $x(t)$ into the set  $\{1,..,\mathcal{N}\}$.
Note that each MOS algorithm can be represented as
\begin{equation} \label{AlModel} 
\hat \nu  = \arg \mathop {\min }\limits_\nu  R\big(\nu, x(t)\big),
\end{equation} 
where $R=R(\nu, x(t))$ is a decision function that completely defines the corresponding MOS algorithm, $\hat \nu$ is the estimate of the number of signals. 
In this paper, we use the decision function $R$ and formalism \eqref{AlModel} to define all the MOS algorithms. For many modern algorithms, the function $R$ is the penalized log-likelihood function, thus we also use the following common representation of the decision function \cite{7465796}: 
\begin{equation} \label{rep_add}
R\left( \nu, x(t) \right)=-\mathit{L}_{\nu}\left(\hat{\bm{\Theta}}_\nu,x(t) \right)+\mathfrak{P}_\nu,
\end{equation}
 where ${L}_\nu={L}_\nu\!\left(\hat{\bm{\Theta}}_\nu,x(t) \right)$ is the logarithm of the likelihood function, $\hat{\bm{\Theta}}_\nu$ denotes the maximum likelihood (ML) estimate of the block vector ${\bm{\Theta}}_\nu$, and $\mathfrak{P}_\nu$ is the so-called penalty term.
 
 Let us consider the problem of estimating the number of modulated sinusoids embedded in AWGN. This problem includes the problem of estimating the number of superimposed sinusoids in AWGN as a special case. Model \eqref{SModel} of the observed data for the problem of estimating the number of modulated sinusoids can be represented as follows:
  \begin{equation} \label{OSModel} 
  	x(t) = \sum\limits_{i = 1}^{\rm{\nu_0 }} {a_{0i} f_i(t){\cos}\!\left( \omega_{0i} t - \varphi_{0i} + \Psi_i(t)\right) } + \sigma_0 n(t).
  \end{equation}
 where $a_i \in \mathbb{R}^1$ is the amplitude, $\omega _i \in \Omega_i=(\omega_{1i},\omega_{2i})$ is the frequency, and ${\varphi _i \in \left[0,2\pi \right]}$ is the phase shift of the $i$-th signal, $f_i(t)$, $\Psi_i(t)$ denote the amplitude and phase envelope respectively. 
 We assume that amplitudes and phases of the signals in \eqref{OSModel} are unknown, frequencies of these signals are known or unknown or partially (see the definition below) unknown, and the noise level is known or unknown. The amplitude and phase envelopes we assume to be known.
 
 In this paper, we use the traditional definition for the SNR $z_i^2$ of the $i$-th signal in \eqref{OSModel}
 \begin{equation} \label{SNR_def}
 z_i^2=a_{0i}^2/2\sigma_0^2, \, z_i^2,\mathrm{dB}=10\log_{10}\left(  a_{0i}^2/2\sigma_0^2\right) .  
 \end{equation}
 
 Under the above assumptions we can write the log-likelihood function for the observed data \eqref{SModel} as follows
 \begin{equation} \label{LF} 
 	{L}_\nu=-\frac{N_s}{2}\ln\left( 2\pi\sigma^2\right)  -\sum\limits_{t = 1}^{N_s}\frac{\left(x(t)- s_\nu(t,\bm{\Theta}_\nu)\right)^2}{2\sigma^2}.
 \end{equation} 
 
 Firstly, discuss the case of the known noise level. Some terms in \eqref{LF} do not affect the algorithms considered in this paper, thus we can omit these terms and rewrite \eqref{LF} without losing the generality as
 \begin{equation} \label{LFR} 
 	\mathit{L}_\nu =\frac{1}{\sigma^2}\sum\limits_{t = 1}^{N_s}x(t) s_\nu(t,\bm{\Theta}_\nu) -\frac{1}{2\sigma^2}\sum\limits_{t = 1}^{N_s}s_\nu^2(t,\bm{\Theta}_\nu),
 \end{equation} 
 
 Next, suppose that the noise level is unknown. In this case, we should substitute the ML estimate of the noise level instead of ${\sigma^2}$ in \eqref{LF}, $\mathit{L}_{\sigma \nu}=\mathit{L}_\nu\left(\hat{\sigma}^2_{\nu} \right)$
 \begin{equation} \label{Lfull}
 \mathit{L}_{\sigma \nu}\!=\!-\frac{N_s}{2}\ln\!\!\left( \sum\limits_{t = 1}^{N_s}\frac{\left(x(t)- s_\nu(t,\bm{\Theta}_\nu)\right)^2}{N_s/2\pi}\right)  -\frac{N_s}{2}.
 \end{equation} 
 
 Finally, let us use model \eqref{OSModel}, substitute maximum likelihood estimates of the unknown amplitudes and phases to \eqref{LFR} and \eqref{Lfull}, and rewrite the result using the block matrix notation
 \begin{equation} \label{MLFR} 
 	\tilde{\mathit{L}}_{\nu}=\frac{1}{2\sigma_0^2}{\mathbf{X}}_{2\nu}^T {\mathbf{C}}_{2\nu}^{ - 1}{{\mathbf{X}}_{2\nu}}, 
 \end{equation}
 \begin{equation} \label{LFrSig} 
 	\tilde{\mathit{L}}_{\sigma \nu}\! =\!-\frac{N_s}{2}\ln\!\!\left(\!\frac{\sum\limits_{t = 1}^{N_s}{\left(x^2(t)\right)}-{\mathbf{X}}_{2\nu}^T {\mathbf{C}}_{2\nu}^{ - 1}{{\mathbf{X}}_{2\nu}}}{N_s/ 2\pi}\!\right)  \!\!-\!\frac{N_s}{2},
 \end{equation}
 where the superscript $T$ denotes the transpose, ${{\mathbf{X}}_{n}} = \left( {{X_i}} \right)_{i = 1}^{n} $, ${\mathbf{{C}}_{n}} = ({C}_{i,j})_{i,j=1}^{n} $, $k = \lceil i/2 \rceil$, $m = \lceil j/2 \rceil$,
 {\small
 \[{X_i} \!=\! \left\{ \!\!\!\! \begin{array}{l}
 {X_{ck}}, \mbox{ if } i \mbox{ is even,}\\
 {X_{sk}}, \mbox{ if } i \mbox{ is odd};
 \end{array} \right.\! {C_{i,j}} \!=\! \left\{ \!\! \begin{array}{l}
 {C_{ckm}},   \mbox{ if } i \mbox{ is even} \mbox{ and } j \mbox{ is even},\\
 {C_{skm}},   \mbox{ if } i \mbox{ is odd } \mbox{ and } j \mbox{ is odd},\\
 {C_{cskm}}, \mbox{if } i \mbox{ is even} \mbox{ and } j \mbox{ is odd},\\
 {C_{sckm}}, \mbox{if } i \mbox{ is odd } \mbox{ and } j \mbox{ is even};
 \end{array} \right.\] 
 \begin{equation} \label{datvec}
 {\left\lbrace \!\!\!\! \begin{array}{l}
 	{X_{ck}}\\
 	{X_{sk}}
 	\end{array}\!\!\!\! \right\rbrace } = \sum\limits_{t=1}^{N_s} x(t){f_k}(t){\left\lbrace \!\!\!\! \begin{array}{l}
 		{\cos}\\
 		{\sin}
 		\end{array}\!\!\!\! \right\rbrace }\! ({\omega _k}t + {\Psi _k}(t)),
 \end{equation} 
 \begin{equation*}
 {\left\lbrace \!\!\!\! \begin{array}{l}
 	{C_{ckm}}\\
 	{C_{skm}}
 	\end{array}\!\!\!\! \right\rbrace }\!\!=\!\!\sum\limits_{t=1}^{N_s} {{f_k}(t){f_m}(t){\left\lbrace \!\!\!\! \begin{array}{l}
 			{\cos}\\
 			{\sin}
 			\end{array}\!\!\!\! \right\rbrace }\!({\omega _k}t\! +\! {\Psi _k}(t)){\left\lbrace \!\!\!\! \begin{array}{l}
 					{\cos}\\
 					{\sin}
 					\end{array}\!\!\!\! \right\rbrace }\!({\omega _m}t \!+\! {\Psi _m}(t))},\\
 \end{equation*}
 \begin{equation*}
 {\left\lbrace \!\!\!\! \begin{array}{l}
 	{C_{cskm}}\\
 	{C_{sckm}}
 	\end{array}\!\!\!\! \right\rbrace }\!\!=\!\!\sum\limits_{t=1}^{N_s} {{f_k}(t){f_m}(t){\left\lbrace \!\!\!\! \begin{array}{l}
 			{\cos}\\
 			{\sin}
 			\end{array}\!\!\!\! \right\rbrace }\!({\omega _k}t \! +\! {\Psi _k}(t)){\left\lbrace \!\!\!\! \begin{array}{l}
 					{\sin}\\
 					{\cos}
 					\end{array}\!\!\!\! \right\rbrace }\!({\omega _m}t \!+\! {\Psi _m}(t))}.
 \end{equation*}
 }
 
 In the case of unknown or partially unknown frequencies, we need to replace the frequencies in \eqref{MLFR} by their estimates. 
 In this paper, we study two different (in terms of frequency estimation) approaches to the design of MOS algorithms.
 
 For the first approach  (ML approach), we should substitute ML estimates of frequencies $\hat{\omega}_i$ into \eqref{MLFR} instead of the unknown frequencies $\omega_i$:
 $\check{\mathit{L}}_{\nu}=\tilde{\mathit{L}}_{\nu}\left(\hat{\bm{\omega}}_\nu, x(t) \right)$ and $\check{\mathit{L}}_{\sigma \nu}=\tilde{\mathit{L}}_{\sigma \nu}\left(\hat{\bm{\omega}}_\nu, x(t) \right),$
 where $\hat{\bm{\omega}}_\nu=\left( \hat{\omega}_i\right)_{i=1}^\nu$. The ML approach is universal and well-known.
 
 For the second approach (the quasilikelihood (QL) approach, see \cite{5762644, White82}) we should substitute arbitrary values $\omega_i^* \in \Omega_i$ of frequencies into \eqref{MLFR} instead of the unknown frequencies $\omega_i$:
 $\tilde{\mathit{L}}_{\nu}^*=\,\tilde{\mathit{L}}_{\nu}\left(\bm{\omega}_\nu^*, x(t) \right)$ and $\tilde{\mathit{L}}_{\sigma \nu}^*=\,\tilde{\mathit{L}}_{\sigma \nu}\left(\bm{\omega}_\nu^*, x(t) \right)$. This simple estimate of frequencies we call a blind (BL) estimate. Moreover, if one uses BL estimates instead of the ML ones in some MOS algorithms we call such MOS algorithms QL algorithms (see, for example, \cite{TrifonovKharin5}, \cite{TrifonovKharin6}).
 
 The BL estimates and the QL approach can be useful for an estimation problem in which there are some {\it a priori} known domains $\Omega_i$ containing unknown frequencies $\omega_i$ and these domains are {\it small enough}. We call such an estimation problem a problem of estimating the number of modulated sinusoids with {\it partially} unknown frequencies.   
 
 \section{Algorithms Design} \label{algdes}
{\it Definition}.  A non-adaptive MOS algorithm is the MOS algorithm whose decision function has the following structure: 
 $R=-{L}_\nu +\mathfrak{K}\cdot\nu,$
where $\mathfrak{K}$ is some term that does not depend on the observed data.

Now define the decision function of the non-adaptive MOS algorithm based on the generalized information criterion (GIC) \cite{Stoica2004, Mariani2015}
\begin{equation} \label{GIC_R} 
R_{\scriptscriptstyle GIC}=-{L}_\nu +\varUpsilon\kappa\nu,
\end{equation} 
here $\kappa= \phi(\nu)/\nu$, $\phi(\nu)$ denotes the number of independently adjusted parameters within the model (see, \cite{AIC}), and $\varUpsilon$ is a constant or a function of $N_s$, $\kappa$ or the other parameters. The generalized information criterion includes several different criteria (AIC, BIC, MDL, criterion in \cite{Nadler2011}, etc.). 

Non-adaptive algorithms have some flaws. In particular, non-adaptive algorithms are high-SNR inconsistent (see \cite{7465796} or Theorem \ref{negG} in this paper). 
{\it Adaptive} MOS algorithms can solve the inconsistency flaw and thus they can be more efficient in many scenarios. 
Penalty functions of the adaptive algorithms do not have any fixed structure.
In \cite{7465796} and other sources one can find a wide range of adaptive MOS algorithms based on the various approaches. Here we provide algorithms based on the EEF \cite{EEF}, g-MDL \cite{gmdl}, NMDL \cite{nmdl}, and RRT \cite{RRT} criteria as examples (note that, the RRT-based algorithm is used only in the case of the unknown noise level).

As mentioned in Section \ref{myintro}, the error probability (or equivalent measures) is a very important performance measure of MOS algorithms for estimating the number of superimposed sinusoids. In practical applications, the minimum error probability criterion is also very important for most signal processing systems because the errors of the used MOS algorithm usually affect the performance of the system as a whole (for example, \cite{RadarApp1}, \cite{RadarApp2},\cite{CommApp1},\cite{RAapp}, \cite{SpectralApp1}, \cite{MOSapp}). However, almost all modern MOS algorithms (both adaptive and non-adaptive) are not based on the minimum error probability criterion or some related criteria (for example, \cite{AIC}, \cite{BIC}, \cite{MDL}, \cite{MAP}, \cite{EEF}, \cite{PAL}, \cite{gmdl}, \cite{nmdl}). 
This is primarily because the following result.

\begin{theorem} \label{no_ump}
There exists no MOS algorithm that uniformly minimizes the error probability $p_e$ for all possible true values of the unknown signal parameters $\bm{\Theta}_{\nu_0}$.
\end{theorem}

\begin{IEEEproof}
The estimation of the true number of signals $\nu_0 \in \{1, \ldots, \mathcal{N}\}$ can be formulated as a multiple composite hypothesis testing problem. Under each hypothesis $H_\nu$ ($\nu_0 = \nu$), the parameter vector $\bm{\Theta}_\nu$ is an unknown nuisance parameter. Moreover, the dimensionality of the parameter space varies across the hypotheses. On the one hand, according to statistical detection theory \cite{Lehmann2005, Kay1998}, a Uniformly Most Powerful (UMP) test does not exist for composite hypotheses with multi-dimensional unknown nuisance parameters. On the other hand, UMP test is required to uniformly minimize the error probability over the entire parameter space. Consequently, an optimal MOS algorithm in the sense of the minimum error probability cannot be designed.
\end{IEEEproof}

Due to this fundamental limitation, existing MOS criteria do not minimize the exact error probability. Taking this into account, we present a new approach to the design of MOS algorithms that is \textit{partially} based on the minimum error probability criterion. Let us provide the following step-by-step description of this approach.

{\it Firstly}, one should design a parametric family of MOS algorithms, i.e., it is needed to design a structure of the MOS algorithm that is defined up to some vector of parameters (the tuning parameters). This structure should generate consistent MOS algorithms at least on some subset (the consistent subset) of the tuning parameters.
Note that each MOS algorithm from the designed parametric family corresponds to a single value of the tuning parameters vector.
{\it Secondly}, it is needed to obtain the error probability of each consistent MOS algorithm from the designed family, i.e., we need to obtain the error probability of the corresponding MOS algorithm for each value of the tuning parameters vector from the consistent subset.  
{\it Finally}, one should find the optimum value of the tuning parameters vector in the sense of minimum error probability. 
The criteria based on the described approach are called the parametric minimum error probability (PMEP) criteria.  
 
In this paper, we focus on three examples of applications of the PMEP approach and three corresponding penalty terms. Two of these terms were first proposed and used for the particular problems in \cite{TrifonovKharin1, TrifonovKharin2, TrifonovKharin3, TrifonovKharin4}.
The first penalty term is called an invariant random penalty term, see \cite[Eq. (30)]{TrifonovKharin1}. Using representation \eqref{AlModel} the MOS algorithm with the invariant random penalty term (PMEP-IR algorithm) can be defined by the following decision function
\begin{equation} \label{RIT} 
R_{\scriptscriptstyle IR}= -{L}_\nu+\kappa_{\scriptscriptstyle IR} \nu \!\! \mathop {\max }\limits_{i \in \left\lbrace 1,\ldots, \mathcal{N} \right\rbrace}\left({{L}_i  -{L}_{i-1}}\right), 
\end{equation} 
where ${L}_0$ is the logarithm of the likelihood function in the case of $\nu=0$ (i.e., under the assumption that there is no signal in the observed data), ${\kappa_{\scriptscriptstyle IR}>0}$ is a tuning parameter.   

The second penalty term  is called an inverse penalty term, see \cite[Eq.31]{TrifonovKharin1} . The MOS algorithm with the inverse penalty term (PMEP-I algorithm) can be defined as
\begin{equation} \label{IT} 
R_{\scriptscriptstyle I}= -\left( {L}_\nu-L_0\right)^{\kappa_{\scriptscriptstyle I}}\!\!/ \nu,
\end{equation} 
where ${\kappa_{\scriptscriptstyle I}>0}$ is a tuning parameter.

Finally, in this paper we present for the first time the complex penalty term (PMEP-C algorithm)
\begin{equation} \label{complexAlg} 
 R_{C}= -{L}_\nu +\kappa_{1c}\nu+ \kappa_{2c} \nu \left( \mathop {\max }\limits_{i \in \left\lbrace 1,\ldots, \mathcal{N} \right\rbrace}\left({{L}_i  -{L}_{i-1}}\right)\right)^{\kappa_{3c}} ,
\end{equation} 
where ${\kappa_{1c}}$, ${\kappa_{2c}}$, and ${\kappa_{3c}}$ are tuning parameters. Note that, parameters ${\kappa_{1c}}$ and ${\kappa_{2c}}$ serve to minimize error probability, and parameter ${\kappa_{3c}}$ serves to provide a balance between the local $\kappa_{1c}\nu$ and global $\kappa_{2c} \nu \!\left(...\right)^{\kappa_{3c}}$ components of the complex penalty term for different values of SNR.

Now let us use \eqref{RIT}, \eqref{IT}, and \eqref{complexAlg} with the PMEP criterion. Following the PMEP criterion, we should first find the subsets of the tuning parameters ${\kappa_{\scriptscriptstyle IR}}$, ${\kappa_{\scriptscriptstyle I}}$, and $\{{\kappa_{1c}}, {\kappa_{2c}},{\kappa_{3c}}\}$ such that algorithms \eqref{RIT}, \eqref{IT}, and \eqref{complexAlg} are consistent (please, see Section \ref{consistency}). Secondly, we should calculate the error probabilities of algorithms \eqref{RIT}, \eqref{IT}, \eqref{complexAlg} for each value of the corresponding tuning parameters ${\kappa_{\scriptscriptstyle IR}}$, ${\kappa_{\scriptscriptstyle I}}$,$\{{\kappa_{1c}}, {\kappa_{2c}},{\kappa_{3c}}\}$ from the found consistent subsets (please, see Section \ref{aep_calc} taking into account the results from Section \ref{consistency}). Finally, using the consistent subsets and calculated error probabilities, it is necessary to find optimal in the sense of minimum error probability values of the tuning parameters ${\kappa_{\scriptscriptstyle IR}}$, ${\kappa_{\scriptscriptstyle I}}$, and $\{{\kappa_{1c}}, {\kappa_{2c}},{\kappa_{3c}}\}$ for the problems under consideration (please, see Section \ref{sims}).

It will be shown in Section \ref{PMEPfuture} that algorithms \eqref{RIT}, \eqref{IT} have some significant limitations. However, it is also important to note that these algorithms are only special cases of the PMEP approach, and we consider them here only as examples of the implementation of the PMEP approach.

\section{Consistency and performance analysis} \label{perfan}
\subsection{Properties of likelihood functions} \label{PreRes}
First of all, let us provide a general framework that is needed for the further consistency and performance analysis of the MOS algorithms. Note that, the results below are obtained using the results given in Appendix \ref{prelim} (including a new representation for the Gram-Schmidt orthogonalization, see Theorem \ref{NewfGS} in Appendix \ref{newNIGS})

Represent \eqref{MLFR} as
\begin{equation} \label{Main_Rep}
\tilde{\mathit{L}}_{\nu}\!=\!\sum_{i=1}^{\nu}\left(\tilde{\mathit{L}}_{i}-\tilde{\mathit{L}}_{i-1} \right)=\frac{1}{2}\sum_{i=1}^{\nu}\left( {l_{si}^2+l_{ci}^2}\right)=\frac{1}{2}\sum_{i=1}^{\nu} {l_i^2},
\end{equation}
where $\tilde{\mathit{L}}_{0}=0$,  $l_{s1}^2=\frac{1}{\sigma_0^2}{\mathbf{X}}_{1}^T {\mathbf{C}}_{1}^{ - 1}{{\mathbf{X}}_{1}}$, $l_i^2=l_{si}^2+l_{ci}^2$, and
\begin{equation} \label{Main_l}
\begin{split}
l_{ci}^2 &=\frac{1}{\sigma_0^2}\left( {\mathbf{X}}_{2i}^T {\mathbf{C}}_{2i}^{ - 1}{{\mathbf{X}}_{2i}} - {\mathbf{X}}_{2i-1}^T {\mathbf{C}}_{2i-1}^{ - 1}{{\mathbf{X}}_{2i-1}}\right) , \\
l_{si}^2 &=\frac{1}{\sigma_0^2}\left( {\mathbf{X}}_{2i-1}^T {\mathbf{C}}_{2i-1}^{ - 1}{{\mathbf{X}}_{2i-1}} - {\mathbf{X}}_{2i-2}^T {\mathbf{C}}_{2i-2}^{ - 1}{{\mathbf{X}}_{2i-2}}\right) . 
\end{split}
\end{equation}

Expressions for random variables $\{l_{ci}\}_{i=1}^{\mathcal{N}}$ and $\{l_{si}\}_{i=1}^{\mathcal{N}}$ (up to sign) in \eqref{Main_l} can be obtained in the same way as in \cite[Eq. 37]{TrifonovKharin1} 
\begin{equation} \label{Main_l_details}
\begin{split}
l_{ci} &=\frac{\gamma_{2i}}{\sigma_0}\left( {X}_{2i} - {\mathbf{X}}_{2i-1}^ T {\mathbf{C}}_{2i-1}^{- 1}{\mathbf{r}}_{2i}\right)/\sqrt{\mathfrak{S}_{2i}}, \\
l_{si} &=\frac{\gamma_{2i-1}}{\sigma_0}\left( {X}_{2i-1} - {\mathbf{X}}_{2i-2}^ T {\mathbf{C}}_{2i-2}^{- 1}{\mathbf{r}}_{2i-1}\right)/\sqrt{\mathfrak{S}_{2i-1}},
\end{split}
\end{equation}
where for all $i$:\\ 
$\mathfrak{S}_i = {C_{i,i}} - {\mathbf{r}}_i^ T \mathbf{C}_{i - 1}^{ - 1}{\mathbf{r}}_i$, ${{\mathbf{r}}_i}=\left( {{C}_{1,i}, \ldots ,{C}_{i-1,i}} \right)^T$, $\gamma_{i}=\pm 1$. 

From \eqref{Main_l_details} and the definition of the random variables $\{{X_i}\}$ one can conclude that random variables $\{l_{ci}, l_{si}\}$ are Gaussian as the linear combinations of the Gaussian random variables. 
Note that random variables $\{l_{ci}\}_{i=1}^{\mathcal{N}}$, $\{l_{si}\}_{i=1}^{\mathcal{N}}$ can be considered as random fields $\{l_{ci}\left(\bm{\omega}_i\right)\}_{i=1}^{\mathcal{N}}$, $\{l_{si}\left(\bm{\omega}_i\right)\}_{i=1}^{\mathcal{N}}$, where ${\bm{\omega}_{i }\!=\!\left(\omega_j\right)_{j=1}^i}$. 
Theorem \ref{randOrh} (see Appendix \ref{prelprob}) and the above results lead to the following consequences.
\begin{theorem} \label{fieldOrh}
The random fields from the following set
$$ \left\{ \left\{l_{ci}\left(\bm{\omega}_i\right) \right\}_{i=1}^{\mathcal{N}} \cup \left\{l_{si}\left(\bm{\omega}_i\right) \right\}_{i=1}^{\mathcal{N}} \right\} $$
are independent and unit-variance Gaussian random fields.
\end{theorem}
Also, if one suppose that $\bm{\omega}_{i}=\bm{\omega}_{0i }$ for all $i$, then the following result can be obtained using Theorem \ref{zeroE} (see Appendix \ref{prelprob})
\begin{equation} \label{prop2}
\mathrm{E}(l_{ci}\left(\bm{\omega}_{0i}\right))=\mathrm{E}(l_{si}\left(\bm{\omega}_{0i}\right))=0,\,\,\mathrm{for\,all}\,\,i>\nu_0.
\end{equation} 

Next, rewrite ${\tilde{\mathit{L}}}_{\nu}^*$  and $\check{\mathit{L}}_{\nu}$ by analogy with \eqref{Main_Rep}:
\begin{equation} \label{Main_Rep_K}
\tilde{\mathit{L}}_{\nu}^*=\frac{1}{2}\sum_{i=1}^{\nu}\left( {l_{si}^{*2}+l_{ci}^{*2}}\right),
\check{\mathit{L}}_{\nu}=\frac{1}{2}\sum_{i=1}^{\nu}\left( {\check{l}_{si}^2+\check{l}_{ci}^2}\right),
\end{equation}
where random variables $l_{si}^*$ and $l_{ci}^*$ in \eqref{Main_Rep_K} are obtained from $l_{si}$ and $l_{ci}$ in  \eqref{Main_Rep} by the replacement of $\omega_i$ by $\omega_i^*$; $\check{l}_{si}$ and $\check{l}_{ci}$ in \eqref{Main_Rep_K} are obtained from $l_{si}$ and $l_{ci}$ \eqref{Main_Rep} by the replacement of $\omega_i$ by $\hat{\omega}_i$. 

Taking into account Theorem \ref{fieldOrh} we can conclude that $ \left\{ \left\{l_{ci}^* \right\}_{i=1}^{\mathcal{N}} \cup \left\{l_{si}^* \right\}_{i=1}^{\mathcal{N}} \right\} $ are independent unit-variance Gaussian random variables for all $\left\lbrace \bm{\omega}_i^* \right\rbrace$.  
Thus, cumulative distribution function (CDF) $F_{i}^*(x)$ and probability density function (PDF) $W_{i}^*(x)$ of $l_{ci}^{*2}+l_{si}^{*2}$ can be represented as   
\begin{equation} \label{CDF1}
F_{i}^*\!=\! F_{NSQ}\!\left(x, d_{ci}^{*2} \!+\! d_{si}^{*2} \right) \!\!,
W_{i}^*\!=\! W_{NSQ}\!\left(x, d_{ci}^{*2} \!+\! d_{si}^{*2} \right),
\end{equation}
where $F_{NSQ}(x, \lambda)$ denotes CDF of the noncentral chi-square random variable with two degrees of freedom and noncentrality parameter $\lambda$; $W_{NSQ}(x, \lambda)$ denotes PDF of this random variable and $d_{ci}^{*}=\mathrm{E}\left( l_{ci}^{*}\right)$, $d_{si}^{*}=\mathrm{E}\left( l_{si}^{*}\right)$. In particular, statistical properties of random variables $ \left\{ \left\{{l}_{ci}^2 \right\}_{i=1}^{\mathcal{N}} \cup \left\{{l}_{si}^2 \right\}_{i=1}^{\mathcal{N}} \right\} $ in the case of known frequencies can be obtained from \eqref{CDF1} if we put $\bm{\omega}_{i}^*=\bm{\omega}_{0i }$ for all $i$.

Now study statistical properties of random variables $ \left\{ \left\{\check{l}_{ci}^2 \right\}_{i=1}^{\mathcal{N}} \cup \left\{\check{l}_{si}^2 \right\}_{i=1}^{\mathcal{N}} \right\} $ for the ML method.
In this case, we assume that the signals in \eqref{OSModel} are orthogonal, i.e.,
\begin{equation} \label{orth}
\sum\limits_{t = 1}^{N_s} {s_i}(t,{{\bm{\theta}}_{i}}){s_j}(t,{{\bm{\theta}}_{j}})= \left\lbrace 
{\begin{aligned}{E_i \,\, \mathrm{if} \,\, i=j,} \\ {0\,\, \mathrm{if} \,\, i \ne j,} \end{aligned}}
\right.
\end{equation}
in terms of \eqref{SModel}. Also, we assume that the energies of signals $E_i$ do not depend on $\omega_i$ and $\varphi_i$. 

Taking into account condition \eqref{orth}, Theorem \ref{fieldOrh}, and results from \cite{TrifonovKharin4} one can conclude that random variables $ \left\{ \left\{\check{l}_{ci}^2 \right\}_{i=1}^{\mathcal{N} } \cup \left\{\check{l}_{si}^2 \right\}_{i=1}^{\mathcal{N} } \right\} $ are independent. In practical scenarios with finite sample sizes $N_s$, superimposed sinusoidal signals are rarely orthogonal. Let us relax condition \eqref{orth} and assume instead that the true frequencies of the signals are well-separated, i.e., their frequency difference exceeds the Rayleigh resolution limit ($|\omega_{0i} - \omega_{0j}| \gg 2\pi/N_s$ for $i \ne j$). Under this condition, the signals are asymptotically orthogonal. Note that, the off-diagonal elements of the matrix ${\mathbf{C}}_{n}$ are bounded as $\mathcal{O}(1)$, whereas the diagonal elements grow as $\mathcal{O}(N_s)$. Consequently, the random variables $ \left\{ \left\{\check{l}_{ci}^2 \right\}_{i=1}^{\mathcal{N} } \cup \left\{\check{l}_{si}^2 \right\}_{i=1}^{\mathcal{N} } \right\} $ become asymptotically independent.

Using Theorem \ref{fieldOrh}, one can obtain that $l_{ci}^2(\omega) + l_{si}^2(\omega)$ follows a chi-square distribution with two degrees of freedom. Next, using the theory of extreme values of random processes (see, e.g., \cite{TrifonovKharin4}), we conclude that the cumulative distribution function of $\check{l}_{ci}^2+\check{l}_{si}^2$ converges to the following distribution:
\begin{equation} \label{CDF_Slag}
\check{F}_i(x) \approx \begin{cases} 
\Phi\left( \frac{x - N_s z_i^2}{2\sqrt{N_s}z_i} \right) \exp\left( - \frac{\xi_i}{\pi} \exp\left( - \frac{x}{2} \right) \right), & \text{if } i \le \nu_0, \\ 
\exp\left( - \frac{\xi_i}{\pi} \exp\left( - \frac{x}{2} \right) \right), & \text{if } i > \nu_0,
\end{cases}
\end{equation}
here $\xi_i = (\omega_{2i} - \omega_{1i}) \sqrt{ \frac{\sum_{t=1}^{N_s} t^2 f_i^2(t)}{\sum_{t=1}^{N_s} f_i^2(t)} - \left( \frac{\sum_{t=1}^{N_s} t f_i^2(t)}{\sum_{t=1}^{N_s} f_i^2(t)} \right)^2 }.$ 

Moreover, as the SNR grows large ($z_i \to \infty$), this distribution can be approximated as:
\begin{equation} \label{CDF_Simplified}
\check{F}_i(x) \approx \begin{cases} 
\Phi\left( \frac{x - N_s z_i^2}{2\sqrt{N_s}z_i} \right), & \text{if } i \le \nu_0, \\ 
\exp\left( -\frac{\xi_i}{\pi} e^{-x/2} \right), & \text{if } i > \nu_0. 
\end{cases}
\end{equation}
Based on the obtained distributions \eqref{CDF_Slag} and \eqref{CDF_Simplified}, we can establish the following result.

\begin{theorem} \label{Eln_th}
For $i > \nu_0$ and $N_s \to \infty$:
\begin{equation} \label{EV_asymptotic}
\mathrm{E}(\check{l}_{ci}^2 + \check{l}_{si}^2) = 2 \ln N_s + \mathcal{O}(1).
\end{equation}
\end{theorem}

\begin{IEEEproof}
For $i > \nu_0$, the CDF \eqref{CDF_Simplified} is the Gumbel distribution $F(x) = \exp(-A e^{-x/B})$ with $A = \xi_i/\pi$ and $B = 2$. The mean of this distribution is $B \ln A + B\gamma_{\text{EM}}$, where $\gamma_{\text{EM}}$ is the Euler-Mascheroni constant. It implies $\mathrm{E}(\check{l}_{ci}^2 + \check{l}_{si}^2) = 2 \ln(\xi_i/\pi) + 2\gamma_{\text{EM}}$. Taking into account $\xi_i = \mathcal{O}(N_s)$, one can conclude that $\ln \xi_i = \ln N_s + \mathcal{O}(1)$.
\end{IEEEproof}

Let us proceed to the case of unknown noise variance. Represent ${\tilde{\mathit{L}}}_{\sigma \nu}^*$  and $\check{\mathit{L}}_{\sigma \nu}$ by analogy with \eqref{Main_Rep}:
\begin{equation} \label{Main_Rep_Ksigma}
\begin{aligned}
\tilde{\mathit{L}}_{\sigma \nu}^* - \tilde{\mathit{L}}_{\sigma 0}^* &= -\frac{N_s}{2}\sum_{i=1}^{\nu} \ln\left( 1 - l_{\sigma i}^{*2}\right), \\
\check{\mathit{L}}_{\sigma \nu} - \check{\mathit{L}}_{\sigma 0} &= -\frac{N_s}{2}\sum_{i=1}^{\nu} \ln\left( 1 - \check{l}_{\sigma i}^{2}\right).
\end{aligned}
\end{equation}
\begin{theorem} \label{unknownNoiseOrh}
1) The variables $\{l_{\sigma i}^{*2}\}_{i > \nu_0}$ are independent.
2) All random variables from the set $\{l_{\sigma i}^{*2}\}_{i > \nu_0}$ are independent of  $\{l_{\sigma i}^{*2}\}_{i \le \nu_0}$.
3) For $i > \nu_0$, the CDF $F_{\sigma i}^*(x)$ of $ l_{\sigma i}^{*2}$ is the Beta distribution of the first kind:
\begin{equation} \label{Beta_CDF}
F_{\sigma i}^*(x) = I_x\left(1, \frac{N_s - 2i}{2}\right) = 1 - (1 - x)^{\frac{N_s - 2i}{2}},
\end{equation}
for $0 \le x \le 1$, where $I_x(\cdot, \cdot)$ denotes the regularized incomplete Beta function.
4) For $i = \nu_0$, the CDF $F_{\sigma \nu_0}^*(x)$ is the singly noncentral Beta distribution of the first kind with the non-centrality parameter $\lambda_{\nu_0} = d_{c\nu_0}^{*2} + d_{s\nu_0}^{*2}$.
5) For $i > \nu_0$, the CDF $F_{\Delta L, i}^*(v)$ of $\Delta \tilde{\mathit{L}}_{\sigma i}^* $ is the exponential distribution:
\begin{equation} \label{Exp_CDF}
F_{\Delta L, i}^*(v) = 1 - \exp\left( -v \frac{N_s - 2i}{N_s} \right), \quad v \ge 0.
\end{equation}
\end{theorem}
\begin{IEEEproof}
Please see Appendix~\ref{proofUNO}.
\end{IEEEproof}

Finally, suppose that the signal frequencies and the noise level are both unknown. As the number of samples tends to infinity ($N_s \to \infty$), the noise level estimate in the denominator converges in probability to the true noise level. Consequently, the distributions converge to those derived under the condition of a known noise level. Specifically, for $i > \nu_0$, the CDF converges to the Gumbel distribution (see \eqref{CDF_Slag} and \eqref{CDF_Simplified}). For $i \le \nu_0$, as the SNR tends to infinity ($z_i \to \infty$), the CDF converges to the Gaussian CDF $\Phi(\cdot)$ presented in \eqref{CDF_Simplified}. Therefore, the asymptotic distribution functions $\check{F}_i(x)$ and $\check{W}_i(x)$ can be applied to evaluate the error probabilities in the case of an unknown noise level.

For brevity, in this paper we shall assume that $V_i = {l_{si}^{*2}+l_{ci}^{*2}}$, $F_i(x)=F_i^*(x)$, $W_i(x)=W_i^*(x)$ (for the known noise level case) or $V_i = {l_{\sigma si}^{*2}+l_{\sigma ci}^{*2}}$, $F_i(x)=F_{\sigma i}^*(x)$, $W_i(x)=W_{\sigma i}^*(x)$ (for the unknown noise level case) if we study the QL design approach or that $V_i = {\check{l}_{si}^2+\check{l}_{ci}^2}$, $F_i(x)=\check{F}_i(x)$, $W_i(x)=\check{W}_i(x)$ (for the known noise level case) or $V_i = {\check{l}_{\sigma si}^2+\check{l}_{\sigma ci}^2}$, $F_i(x)=\check{F}_{\sigma i}(x)$, $W_i(x)=\check{W}_{\sigma i}(x)$ (for the unknown noise level case) if we study the ML design approach or that $V_i = {l_{si}^{2}+l_{ci}^{2}}$ if we study the general case.

\subsection{Consistency}  \label{consistency}
Now find the ranges of the values of tuning parameters $\kappa_{\scriptscriptstyle IR}$ and $\kappa_{\scriptscriptstyle I}$ which provide the consistency of MOS algorithms \eqref{RIT}  and \eqref{IT}, respectively.
Let us represent SNRs \eqref{SNR_def} of all signals as $z_i=z_bz_{ni}$.  Next, we can write a definition of the SNR-consistency. In terms of \eqref{AlModel} this definition can be written as
\begin{equation} \label{cons} 
\Pr\!\left( \bigcap_{ \substack{\scriptstyle \nu \ne {\nu _0}}} \!\! {\left\lbrace R\big( {\nu _0},x(t)\big)  - R\big( \nu,x(t)\big) \!\! <0\right\rbrace} \! \right) \xrightarrow[z_{b} \to \infty]{} 1.
\end{equation}
Note that the concept of SNR-consistency \eqref{cons} generalizes the same concept in \cite{7465796}. 

Suppose that the noise level is {\it known}. In the case of known frequencies and in the case of the QL design approach to the estimation of unknown frequencies, using representation \eqref{Main_Rep}  we can rewrite \eqref{cons} as $\Pr\left(\mathfrak{C} \right) \xrightarrow[z_{b} \to \infty]{} 1$, where $\mathfrak{C}$ denotes the following intersection of events 
\begin{equation} \label{ir_ce}
\bigcap_{ \substack{\scriptstyle k}}\left\lbrace\sum_{i=\nu_0+1}^{\nu_0+k}{V_{i}}< \kappa_{\scriptscriptstyle IR}k \!\!\! \mathop {\max }\limits_{\scriptscriptstyle  i \in \{ 1,...,{\mathcal{N}}\} } \!\!\!\! V_i \,\,\, <\!\!\! \sum_{i=\nu_0-k+1}^{\nu_0}\!\!\!{V_{i}}\right\rbrace
\end{equation}
for algorithm \eqref{RIT}, or the next intersection of events 
\begin{equation}  \label{i_ce}
\bigcap_{ \substack{\scriptstyle k}}\left\lbrace\!\frac{\sum_{i=\nu_0+1}^{\nu_0+k}{V_{i}}}{ 
  \left(1-\sqrt[{\kappa_{\scriptscriptstyle I}}]{\frac{\nu_0}{\nu_0+k}} \right)} \!<\! \sum_{i=1}^{\nu_0+k}{V_{i}},\, \sum_{i=1}^{\nu_0-k}{V_{i}} \!<\! \frac{\sum_{i=\nu_0-k+1}^{\nu_0}\!{V_{i}}}{ 
    \left(\sqrt[{\kappa_{\scriptscriptstyle I}}]{\frac{\nu_0}{\nu_0-k}}-1 \right)}\!\right\rbrace
\end{equation}
for algorithm \eqref{IT}, respectively. In \eqref{ir_ce} and \eqref{i_ce} the number $1\le k \le \mathcal{N}-\nu_0$ for the left-hand inequalities, and $1 \le k \le \nu_0-1$ for the right-hand inequalities, respectively. 

Taking into account \eqref{Main_l_details} one can conclude that
\begin{equation} \label{dnormz}
d_i^2=\left(d_{ni}^2+\frac{2}{z_b^2}\right)z_b^2,
\end{equation}
here $ d_i^2= \mathrm{E}\left( V_i\right)$, and $d_{ni}^2$ does not depend on $z_b$ for all $i$.

Let us formulate without proof the next simple lemma.
\begin{lemma} \label{ul}
Suppose $\left\{\mathfrak{A}_{i} \right\}_{i=1}^{N} $ is a finite set of events. If all probabilities $\left\lbrace \Pr\left( \mathfrak{A}_{i} \right)\right\rbrace_{i=1}^{N} $ tend to unity, then the probability  $\Pr\!\left( \bigcap_{ \substack{ \scriptstyle i=1}}^{N} \mathfrak{A}_{i} \right) \!\!$ also tends to unity.
\end{lemma}
In the case of known frequencies, using \eqref{prop2}, \eqref{ir_ce}, \eqref{i_ce}, and Lemma \ref{ul} one can find that condition \eqref{cons} is satisfied for algorithms \eqref{RIT} and \eqref{IT} if and only if the following inequalities 
\begin{equation} \label{ce}
 \kappa_{\scriptscriptstyle IR}k \!\!\! \mathop {\max }\limits_{\scriptscriptstyle  i \in \{ 1,...,{\mathcal{N}}\} } \!\!\!\! d_{ni}^2 \!<\!\!\!\!\! \sum_{i=\nu_0-k+1}^{\nu_0}\!\!\!\!\!\!\!\!{d_{ni}^{2}}, \sum_{i=1}^{\nu_0-k}\!{d_{ni}^{2}} \!<\!\frac{\sum_{i=\nu_0-k+1}^{\nu_0}\!{d_{ni}^{2}}}{ 
	\left(\sqrt[{\kappa_{\scriptscriptstyle I}}]{\frac{\nu_0}{\nu_0-k}}-1 \right)},
\end{equation}
hold for all $1 \le k \le \nu_0-1$.

Inequalities \eqref{ce} can be replaced by the following ones
\begin{equation} \label{strong_ce}
\kappa_{\scriptscriptstyle IR}<\rho, \quad \kappa_{\scriptscriptstyle I} \ge 1/\rho, 
\end{equation}
where $d_{min}^2 = \min d_{ni}^2$,  $d_{max}^2 =\max d_{ni}^2 $, and $\rho=d_{min}^2/d_{max}^2$ (please, see more about the parameter $\rho$ in Section \ref{PMEPfuture}). 
Inequalities \eqref{strong_ce} reject some sufficient values of the tuning parameters, but they are quite simpler than inequalities \eqref{ce}.
One can conclude that, if the tuning parameters $\kappa_{\scriptscriptstyle IR}$ and $\kappa_{\scriptscriptstyle I}$ satisfy conditions \eqref{ce} or \eqref{strong_ce}, then the corresponding algorithms are SNR-consistent.

In the case of the QL design approach to the estimation of frequencies, one can find that for $i>\nu_0$ expectations $d_i^2$ can be non-zero (see Section \ref{BLrob}), thus in this case, we get consistency conditions as follows
\begin{equation} \label{ceQL}
\begin{split}
 \sum_{i=\nu_0+1}^{\nu_0+k}{d_{ni}^{2}}< \kappa_{\scriptscriptstyle IR}&k \!\!\! \mathop {\max }\limits_{\scriptscriptstyle  i \in \{ 1,...,{\mathcal{N}}\} } \!\!\!\! d_{ni}^2 \!<\!\!\!\!\! \sum_{i=\nu_0-k+1}^{\nu_0}\!\!\!\!\!\!\!\!{d_{ni}^{2}},\\
 \frac{\sum_{i=\nu_0+1}^{\nu_0+k}{d_{ni}^{2}}}{ 
   \left(1-\sqrt[{\kappa_{\scriptscriptstyle I}}]{\frac{\nu_0}{\nu_0+k}} \right)} <\sum_{i=1}^{\nu_0+k}\!{d_{ni}^{2}},\,&  \sum_{i=1}^{\nu_0-k}\!{d_{ni}^{2}} \!<\!\frac{\sum_{i=\nu_0-k+1}^{\nu_0}\!{d_{ni}^{2}}}{ 
	\left(\sqrt[{\kappa_{\scriptscriptstyle I}}]{\frac{\nu_0}{\nu_0-k}}-1 \right)},
	\end{split}
\end{equation}
for all $1\le k \le \mathcal{N}-\nu_0$ for the left-hand inequalities, and for all $1 \le k \le \nu_0-1$ for the right-hand inequalities, respectively.

Let us study the case of the {\it unknown} noise level. Comparing \eqref{LFR} and \eqref{LFrSig} one can write
\begin{equation} \label{LFrSigDE} 
\begin{split}
 \tilde{L}_{\sigma i}^* - \tilde{L}_{\sigma (i-1)}^*\xrightarrow[z_{b} \to \infty]{} d_{\sigma ni}^2\\
 d_{\sigma ni}^2 =\frac{N_s}{2}\ln\!\!\left(\!\frac{\sum\limits_{j=1}^{\nu_0} {d_{0nj}^2} - \sum\limits_{j=1}^{i-1} {d_{nj}^2}}{\sum\limits_{j=1}^{\nu_0} {d_{0nj}^2} - \sum\limits_{j=1}^{i} {d_{nj}^2}}\!\right),
 \end{split}
                \end{equation}
 where terms $d_{0nj}^2$ can be obtained from $d_{0j}^2 =\mathrm{E}\left(l_{cj}^2\left(\bm{\omega}_{0j}\right)+l_{sj}^2\left(\bm{\omega}_{0j}\right)\right)$ in accordance with formula \eqref{dnormz}.
We can use formulas \eqref{MLFR}, \eqref{LFrSig}, and \eqref{LFrSigDE}, representation \eqref{Main_Rep}, and also the approach to the consistency analysis presented in this subsection (see \eqref{cons}-\eqref{dnormz}) to obtain the consistency conditions for algorithms \eqref{RIT} and \eqref{IT} for the case of known frequencies and unknown noise level:
\begin{equation} \label{strong_ce2}
\kappa_{\scriptscriptstyle IR} < \frac{1}{\mathcal{N} - 1}, \quad \kappa_{\scriptscriptstyle I} > 0. 
\end{equation}
Note that consistency conditions \eqref{strong_ce2} do not contain the parameter $\rho$ or any other similar parameters. Next, if the value of $\mathcal{N}$ grows, then the range of turning parameters where the PMEP-IR algorithm is consistent decreases.
Consistency conditions for the QL design approach can be obtained by replacing in \eqref{ceQL}  terms $d_{ni}^2$ by terms $d_{\sigma ni}^2$ from \eqref{LFrSigDE}. Furthermore, consistency analysis of MOS algorithms proposed in this subsection can be extended to the case when the number of samples tends to infinity. This extension can be done directly if the decision function of the MOS algorithm depends on the observed data and on the number of samples only through the likelihood function \eqref{MLFR} or \eqref{LFrSig}. In particular, for algorithms \eqref{RIT} and \eqref{IT}, one can directly obtain consistency conditions similar to \eqref{ce}, \eqref{strong_ce}, \eqref{ceQL}, \eqref{strong_ce2} and \eqref{ceQL} with \eqref{LFrSigDE} using consistency analysis presented in this subsection.

Finally, let us study the consistency of the MOS algorithms under the ML design approach for estimating unknown frequencies.

\begin{theorem} \label{MLcons}
Under the ML design approach for estimating unknown frequencies, if the tuning parameters satisfy the conditions \eqref{ce}, \eqref{strong_ce}, and \eqref{strong_ce2}, the PMEP MOS algorithms are consistent as the SNR or the number of samples tends to infinity.
\end{theorem}
\begin{IEEEproof}
As $z_b \to \infty$, we get $\hat{\bm{\omega}}_\nu \xrightarrow{\Pr} \bm{\omega}_{0\nu}$. By the continuous mapping and Slutsky's theorems \cite[Theorems 2.3, 2.8]{van1998asymptotic}, $z_b^{-2}V_i(\hat{\bm{\omega}}_\nu) \xrightarrow{\Pr} d_{ni}^2$ for $i \le \nu_0$. For $i > \nu_0$, $V_i(\hat{\bm{\omega}}_\nu) = \mathcal{O}_{\Pr}(1) \implies z_b^{-2}V_i(\hat{\bm{\omega}}_\nu) \xrightarrow{\Pr} 0$. Multiplying the inequalities of \eqref{RIT} and \eqref{IT} by $z_b^{-2}$ as $z_b \to \infty$ yields \eqref{ce}, \eqref{strong_ce}, and \eqref{strong_ce2}.
As $N_s \to \infty$, $N_s^{-1}V_i(\hat{\bm{\omega}}_\nu) \xrightarrow{\Pr} c_i > 0$ for $i \le \nu_0$. By Theorem \ref{Eln_th}, $V_i(\hat{\bm{\omega}}_\nu) = \mathcal{O}_{\Pr}(\ln N_s)$ for $i > \nu_0 \implies N_s^{-1}V_i(\hat{\bm{\omega}}_\nu) \xrightarrow{\Pr} 0$. 
Overestimation ($\hat{\nu} > \nu_0$) in \eqref{RIT} implies $V_{\nu_0+1}(\hat{\bm{\omega}}_\nu) > \kappa_{\scriptscriptstyle IR} \max_{j} V_j(\hat{\bm{\omega}}_\nu)$. Multiplying by $N_s^{-1}$ as $N_s \to \infty$ yields $0 \ge \kappa_{\scriptscriptstyle IR} \max_{j \le \nu_0} c_j$, which is false $\forall \kappa_{\scriptscriptstyle IR} > 0$. 
Underestimation ($\hat{\nu} < \nu_0$) in \eqref{RIT} implies $\kappa_{\scriptscriptstyle IR} \max_{j} V_j(\hat{\bm{\omega}}_\nu) > V_i(\hat{\bm{\omega}}_\nu)$ for some $i \le \nu_0$. Multiplying by $N_s^{-1}$ as $N_s \to \infty$ yields $\kappa_{\scriptscriptstyle IR} \max_{j \le \nu_0} c_j \ge c_i$. Preventing this $\forall i \le \nu_0$ requires $\kappa_{\scriptscriptstyle IR} \max_{j \le \nu_0} c_j < \min_{i \le \nu_0} c_i \implies \kappa_{\scriptscriptstyle IR} < \rho$. An analogous derivation applies to \eqref{IT}.
\end{IEEEproof}

It is important to note that the presented consistency analysis can be extended to a wide range of MOS algorithms. In particular, for SNR-consistency, this extension can be done directly if the decision function of the MOS algorithm depends on the observed data only through the likelihood function (see \eqref{MLFR} or \eqref{LFrSig}) and/or the energy $\sum\limits_{t=1}^{N_s}{x^2(t)}$ of the observed data. For example, decision functions of EEF, g-MDL, NMDL algorithms (see \cite[TABLE II]{7465796}), and decision function \eqref{complexAlg} satisfy this condition.

\subsection{Abridged Error Probability}
First of all, introduce the following definition.
The {\it error probability} of the MOS algorithm $p_e$ is the probability that the estimate obtained by this algorithm does not match with the true number of signals, i.e., $p_e=\Pr(\hat{\nu} \neq \nu_0) $. 
Using representation \eqref{AlModel} we can rewrite the error probability $p_e$ as
\begin{equation} \label{p_e} 
{p_e} = 1 - \Pr\!\left( \bigcap_{ \substack{ \scriptstyle i=1, \scriptstyle i \ne {\nu _0}}}^{\mathcal{N}} {\left\lbrace R\big( {\nu _0}\big)  < R\big( i\big) \right\rbrace}\right) \!\!.
\end{equation}  

Performance analysis by the error probability \eqref{p_e} is quite difficult. Some approximations of the error probability were introduced in recent works (for example, \cite{Nadler2011, Mariani2015, 7465796}). However, these approximations are primarily made for non-adaptive MOS algorithms. Moreover, these approximations do not take into account the general design specifics of the MOS algorithms. 
Namely, in practice applications, the different values of the errors of the MOS algorithms have different effects on the performance of the system as a whole. For example, in the case of $\left| \hat{\nu}-\nu_0\right|=1$  the degradation of the system performance as a whole is usually quite less than in the case of $\left| \hat{\nu}-\nu_0\right| > 1$. Hence, one may choose the next additional condition for the most efficient MOS algorithms 
\begin{equation} \label{cond_1} 
\frac{p\left( \left| \hat{\nu}-\nu_0 \right| >1 \right) }{p\left( \left| \hat{\nu}-\nu_0 \right| =1 \right)}\rightarrow 0,
\end{equation}
as the signal-to-noise ratio (SNR) or the number of samples tends to infinity.
 
Taking into account \eqref{cond_1} we use an abridged error probability \cite{TrifonovKharinRus, TrifonovKharin1, TrifonovKharin2, TrifonovKharin3, TrifonovKharin4, TrifonovKharin5, TrifonovKharin6, KP1} as a universal approximation to the error probability \eqref{p_e}  
\begin{equation} \label{p_a} 
   {p_a} \triangleq 1 - \Pr\!\Big( R\big({\nu _0}\big) < R\big({\nu _0} + 1\big), R\big({\nu _0}\big) < R\big({\nu _0} - 1\big) \Big).
\end{equation}  
 
Let us list some common properties of the abridged error probability. 

P.1. In the case of $ 1<\nu_0<\mathcal{N}$, the abridged error probability \eqref{p_a} is a lower bound of the error probability \eqref{p_e}.

P.2. In the case of $\nu_0=2$, $\mathcal{N}=3$, the error probability \eqref{p_e} is reduced to the abridged error probability \eqref{p_a}.   

P.3. Assume that for any $|i - \nu_0| \ge 2$, the probability of the event $\left\lbrace R(i) < R(\nu_0) \right\rbrace$ decays with the SNR or the number of samples faster than the probability of the events $\left\lbrace R(\nu_0 \pm 1) < R(\nu_0)\right\rbrace$. Thus, as the SNR or the number of samples tends to infinity, we assume that
\begin{equation} \label{assum_1} 
\frac{\Pr\!\left( R(i) < R(\nu_0) \right)}{\Pr\!\left( R(\nu_0 \pm 1) < R(\nu_0) \right)} \rightarrow 0,
\end{equation}
for all $i \in \left\lbrace 1, \ldots, \nu_0 - 2, \nu_0 + 2, \ldots, \mathcal{N}\right\rbrace$. 
Under assumption \eqref{assum_1}, abridged error probability \eqref{p_a} tends to the error probability \eqref{p_e} as the SNR or the number of samples tends to infinity.

Note that, if condition \eqref{assum_1} is not fulfilled, then condition \eqref{cond_1} is also not fulfilled, i.e., the truth of condition \eqref{assum_1} is a necessary property for an efficient algorithm. In practice, even if condition \eqref{assum_1} is not satisfied (for example, for all GIC algorithms condition \eqref{assum_1} is not satisfied), the abridged error probability can be an accurate enough approximation to the error probability (see, for example, Fig. \ref{fig1} and \ref{fig9}). In particular, if instead of zero in conditions \eqref{assum_1} there is a sufficiently small constant, then the approximation with increasing SNR will also be sufficiently accurate. 

\subsection{Theoretical calculation of the abridged error probability} \label{aep_calc}
Using results from Subsection \ref{PreRes} let us calculate the abridged error probability (determined in \eqref{p_a}) for algorithms \eqref{GIC_R}, \eqref{RIT}, \eqref{IT}, and \eqref{complexAlg} in the all cases under consideration (BL/ML frequency estimates, known/unknown noise level). The part of results obtained in this section generalize the ones in \cite[see eq. 46, 51, 53]{TrifonovKharin1} that were obtained for the particular case of the known frequencies and known noise level.

1. Algorithms based on the GIC.
Using \eqref{p_a} and results from Section \ref{PreRes} we obtain the abridged error probability of algorithm \eqref{GIC_R}
\begin{equation} \label{AEP_GIC}
\begin{split}
{p_{a{\scriptscriptstyle GIC}}} &= 1 - \Pr\!\left(V_{\nu_0+1}< 2\varUpsilon\kappa, V_{\nu_0} > 2\varUpsilon\kappa\right)\\
 &= 1 - \Pr\!\left(V_{\nu_0+1}< 2\varUpsilon\kappa\right)\Pr\!\left( V_{\nu_0} > 2\varUpsilon\kappa\right)\\
 &= 1 - F_{\nu_0+1}\left(2\varUpsilon\kappa\right)+F_{\nu_0+1}\left(2\varUpsilon\kappa\right)F_{\nu_0}\left(2\varUpsilon\kappa\right)\!.
\end{split}
\end{equation} 
Note that, if the SNR or $N_s$ tends to infinity, then using \eqref{Main_Rep} and \eqref{Main_l_details} we get
$\Pr\!\left( V_{\nu_0} > 2\varUpsilon\kappa\right) \rightarrow 1$.
Combining this result with \eqref{AEP_GIC} one obtain
\begin{equation} \label{AEP_asimpt2}
{p_{a{\scriptscriptstyle GIC}}} \rightarrow 1 - \Pr\!\left(V_{\nu_0+1}< 2\varUpsilon\kappa\right)= 1 - F_{\nu_0+1}\left(2\varUpsilon\kappa\right).
\end{equation}
If we do not consider the QL design approach, then the right side of \eqref{AEP_asimpt2} does not depend on SNR.

Result \eqref{AEP_GIC} can be used to design new MOS algorithms. Let us provide some examples. Firstly, we fix some arbitrary values of SNRs  $\mathbf{z} = \left(z_i\right)$ and calculate the corresponding values of $\left( d_i^2 \right)$ using \eqref{datvec} and \eqref{Main_l_details}. Next, we find an optimal value of the parameter $\varUpsilon$ (from \eqref{GIC_R}) as

\begin{equation} \label{GIC_min}
\varUpsilon_{opt}\left(\mathbf{z}\right)  = \arg \mathop {\min }\limits_\varUpsilon  {p_{a{\scriptscriptstyle GIC}}}\left(\varUpsilon, \mathbf{z} \right) 
\end{equation}
This approach guarantees that the MOS algorithm \eqref{GIC_R} has the smallest abridged error probability for given SNRs values $\mathbf{z}$. Moreover, if we have some estimation $\hat{\mathbf{z}}$ of the SNRs, then we can use $\hat{\mathbf{z}}$ instead of ${\mathbf{z}}$ in \eqref{GIC_min}. 
 
2. PMEP-IR algorithm.
Firstly, rewrite the abridged error probability \eqref{p_a} for algorithm \eqref{RIT} as
\begin{equation*} 
{p_{a{\scriptscriptstyle IR}}} = 1 - p(V_{{\nu _0} + 1} \!<\! \kappa_{\scriptscriptstyle IR} \!\!\! \mathop {\max }\limits_{i \in \left\lbrace 1,\ldots, \mathcal{N} \right\rbrace }\!\!\!\!\!\!{V_i} ,
 V_{\nu _0} > \kappa_{\scriptscriptstyle IR} \!\!\! \mathop {\max }\limits_{i \in \left\lbrace 1,\ldots, \mathcal{N} \right\rbrace }\!\!\!\!\!\!{V_i}).
\end{equation*} 
Using this formula and the result \cite[Eq. 50]{TrifonovKharin1} we can obtain the following equation for the abridged error probability of algorithm \eqref{RIT}
\begin{equation} \label{p_a_RIT2}
\begin{split}
&p_{a{\scriptscriptstyle IR}} = 1 - \int\limits_0^\infty  {{W_{{\nu _0}}}}\!(x){F_{max}}\!\!\left(\frac{x}{\kappa_{\scriptscriptstyle IR}}\right)\!\!{F_{{\nu _0} + 1}}\!\!\left(x\right)\!\textrm{d}x  \\
&+\! \int\limits_0^\infty \! {{W_{{\nu _0} + 1}}} (x){F_{max}}\!\left(\frac{x}{\kappa_{\scriptscriptstyle IR}}\right)\!\!\!\left(\!F_{\nu _0}\!\!\left(\frac{x}{\kappa_{\scriptscriptstyle IR}} \right)\!\!- {F_{\nu _0}(x)} \!\!\right)\!\textrm{d}x,
\end{split}
\end{equation} 
where $F_{max}$ denotes the CDF of $\mathop {\max }\limits_{ \substack{ \scriptstyle i \in \left\lbrace 1, \ldots, \mathcal{N}\right\rbrace,  \\ \scriptstyle i \ne \nu_0, i \ne \nu_0 + 1.}} \!\!\!{V_i}$, and can be calculated as
\[F_{max} (x) = \prod\limits_{ \substack{ \scriptstyle i = 1,
		\scriptstyle i \ne {\nu _0},i \ne {\nu _0} + 1}}^{{\mathcal{N}}} \!\!\!\!\!\!\!\!\!{{F_i}(x)} .\]

3. PMEP-I algorithm.
The abridged error probability \eqref{p_a} for algorithm \eqref{IT} can be rewritten as 
\begin{equation} \label{p_a_I}
\begin{split}
p_{a{\scriptscriptstyle I}} =& 1 - \Pr\left(V_{\nu _0 + 1}/B -V_{\nu _0}<\sum\limits_{i = 1}^{\nu _0 - 1}V_i <V_{\nu _0}/A \right)\\
=&1 - \!\!\int\limits_0^\infty W_{\nu _0 + 1}(y)\!\!\!\!\int\limits_{\frac{A}{B(A+1)}y}^\infty{\!\!\!\! W_{\nu _0}(x)}\\
&\mspace{80mu}\times{\left[ {F_{\Sigma}(x/A) - F_{\Sigma}\left(y/B - x\right)} \right]\textrm{d}x\textrm{d}y},
\end{split}
\end{equation} 
where $A=\sqrt[\kappa_{\scriptscriptstyle I}]{\frac{\nu _0}{\nu _0 -1}} -1$, $B=\sqrt[\kappa_{\scriptscriptstyle I}]{\frac{\nu _0+1}{\nu _0}} -1$, $F_{\Sigma}$ is the CDF of $\sum_{i=1}^{\nu_0-1}{V_i}$.

4) PMEP-C algorithm. Finally, let us present the abridged error probability for algorithm \eqref{complexAlg}.
 Generalizing the approach used for \eqref{p_a_RIT2}, one can obtain
\begin{equation} \label{p_a_c}
\begin{split}
p_{aC} =& 1 -\!\!\! \int\limits_{x > g(x)} \!\! W_{\nu_0}(x) F_{max}(x) F_{\nu_0+1}\big(g(x)\big) \mathrm{d}x \\
&- \!\!\!\!\!\! \int\limits_{x > g(x)} \!\! W_{max}(x) \Big[ F_{\nu_0}(x) - F_{\nu_0}\big(g(x)\big) \Big] F_{\nu_0+1}\big(g(x)\big) \mathrm{d}x,
\end{split}
\end{equation}
where $g(x) = 2\kappa_{1c} + 2^{1-\kappa_{3c}}\kappa_{2c} x^{\kappa_{3c}}$. Setting $\kappa_{1c}=0$, $\kappa_{3c}=1$, and $\kappa_{2c} = \kappa_{\scriptscriptstyle IR}$ reduces \eqref{p_a_c} to \eqref{p_a_RIT2}.

Formulas \eqref{CDF1}, \eqref{CDF_Slag} with abridged error probability formulas \eqref{AEP_GIC}, \eqref{p_a_RIT2}, \eqref{p_a_I}, and \eqref{p_a_c} allow us to provide a performance analysis of the MOS algorithms based on \eqref{GIC_R}, \eqref{RIT}, \eqref{IT}, and \eqref{complexAlg}.
It is important to note that the presented theoretical performance analysis can be extended to a wide range of MOS algorithms. In particular, one can use presented performance analysis to study the performance of algorithm \eqref{complexAlg}.

\section{Quasilikelihood approach} \label{BLrob}

\subsection{Frequency errors and MOS algorithms robustness} \label{BLROB}
Let us start with the case of known frequencies.
In this case, using result \eqref{AEP_GIC}, statistical properties of random variables $ \left\{ \left\{{l}_{ci}^2 \right\}_{i=1}^{\mathcal{N}} \cup \left\{{l}_{si}^2 \right\}_{i=1}^{\mathcal{N}} \right\} $ (see Section \ref{PreRes}), and property P.1. of the abridged error probability one can find the next theorem.
\begin{theorem} \label{negG}
If $\nu_0<\mathcal{N}$, then the error probability of any non-adaptive MOS algorithm tends to some non-zero constant with the SNR or the number of samples tends to infinity.
\end{theorem}

Suppose that the BL estimates of frequencies $\bm{\omega}_{i}^*$ are used in some non-adaptive MOS algorithm, then the following theorem can be stated. 
\begin{theorem} \label{negnegG}
If $\bm{\omega}_{i}^* \ne \bm{\omega}_{0i }$ and $\nu_0<\mathcal{N}$, then the error probability of the non-adaptive MOS algorithm tends to unity as the SNR or the number of samples tends to infinity.
\end{theorem}
\begin{IEEEproof}
 If the condition $\bm{\omega}_{i}=\bm{\omega}_{0i }$ is not fulfilled, then equalities \eqref{prop2} are also broken. Moreover, in this case, the expectations from \eqref{prop2} depend on the true values of the amplitudes. 
 Indeed, suppose that $\bm{\omega}_{i}=\bm{\omega}_{i}^*$, $\mathrm{E}(l_{ci}^*)$ and $\mathrm{E}(l_{si}^*)$ can be rewritten (up to sign) as follows
\begin{equation} \label{le}
\mathrm{E}\!\left(\!\!\frac{{X}_{i}^* \!-\! {\mathbf{X}}_{i-1}^{*T} {\mathbf{C}}_{i-1}^{**- 1}{\mathbf{r}}_{i}^{**}}{\sigma_0\sqrt{\mathfrak{S}_{i}^{**}}} \!\!\right)\!\!=\!\mathbf{Q}_{0\nu_0}^T \!\!\left(\frac{{\mathbf{r}}_i^*\!-\! \mathbf{C}_{i - 1}^{*T} \mathbf{C}_{i - 1}^{**- 1}{\mathbf{r}}_i^{**}}{\sqrt{\mathfrak{S}_{i}^{**}}} \right)\!\!,
\end{equation}
where ${\mathbf{X}}_{i}^{*}=\left(X_j^*\right)_{j=1}^{i}$, ${{\mathbf{r}}_i^*}=\left({C}_{\alpha,i}^*\right)_{\alpha=1}^{i}$, ${\mathbf{r}}_i^{**}=\left({C}_{\alpha,i}^{**}\right)_{\alpha=1}^{i}$,
$ {\mathbf{C}}_{i}^*=\{{C}_{\alpha,\beta}^*\}_{\alpha,\beta=1}^{i}$, 
${\mathbf{C}}_{i}^{**}=\{{C}_{\alpha,\beta}^{**}\}_{\alpha,\beta=1}^{i}$,
$C_{\alpha,\beta}^*={\mathop{\mathrm{cov}}}\left( {X_\alpha^*,X_\beta}\right) $,   $C_{\alpha,\beta}^{**}={\mathop{\mathrm{cov}}}\left( {X_\alpha^*,X_\beta^*}\right)$, $\mathfrak{S}_i^{**} = {C_{i,i}^{**}} - {\mathbf{r}}_i^{**T} \mathbf{C}_{i - 1}^{** - 1}{\mathbf{r}}_i^{**}$.
Here, the random variables $X_{j}^*$ are obtained from $X_{j}$ in \eqref{MLFR} by replacing the true frequencies $\omega_j$ with their BL estimates $\omega_j^*$. The vector $\mathbf{Q}_{0{\nu_0}}$ contains the true amplitudes and is defined as
\[\left( \mathbf{Q}_{0{\nu_0}}\right)_i  = \left\{ \!\! \begin{array}{l}
\left\{ \!\! \begin{array}{l}
\left( {{a_{0m}}\cos {\varphi _{0m}}}\right) /\sigma_0, \mbox{ if } i \mbox{ is even,}\\
\left( {{a_{0m}}\sin {\varphi _{0m}}}\right) /\sigma_0, \mbox{ if } i \mbox{ is odd};
\end{array} \right., \mbox{ if } i \le \nu_0\\
\,\,\,\,\,\,\,\,\,\,\,\,\,\,\,\,\,\,\,\,\,\,\,\,\,\,\,\, 0, \,\,\,\,\,\,\,\,\,\,\,\,\,\,\,\,\,\,\,\,\,\,\,\,\,\,\,\,\,\,\,\,\,\,\,\,\,\,\,\,\,\,\,\,\,\,\,\,\,\,\,\,\,\, \mbox{ if } i >\nu_0,
\end{array} \right.\]
here $m = \lceil i/2 \rceil$.
Consequently, for any $i>\nu_0$, equation \eqref{le} can be rewritten as
\begin{equation} \label{depen}
\mathrm{E}(l_{ci}^*)= \mathbf{Q}_{0\nu_0}^T \mathbf{k}_{ci}, \quad \mathrm{E}(l_{si}^*)=\mathbf{Q}_{0\nu_0}^T \mathbf{k}_{si},
\end{equation}
where the vectors $\mathbf{k}_{ci}$ and $\mathbf{k}_{si}$ are independent of the true signal amplitudes, the noise level, and the SNR.

Using \eqref{depen}, one concludes that the components of $\mathbf{Q}_{0\nu_0}$ scale linearly with $a_{0i}/\sigma_0$. This implies that for $i>\nu_0$, the expected values $\mathrm{E}(l_{ci}^*)$ and $\mathrm{E}(l_{si}^*)$ grow linearly with $\sqrt{\mathrm{SNR}}$ and $\sqrt{N_s}$. As a result, the non-centrality parameters of the random variables $V_i^* = l_{ci}^{*2} + l_{si}^{*2}$ grow linearly with the SNR and the number of samples. On the other hand, in non-adaptive algorithms, the penalty term $\mathfrak{K} \cdot \nu$ is independent of the SNR. Therefore, as the SNR tends to infinity, $V_i^*$ for $i > \nu_0$  exceeds $\mathfrak{K}$, thus the algorithm selects an overestimated model order. It is important to note that, if a non-adaptive penalty $\mathfrak{K} \cdot \nu$ is designed to grow linearly with $N_s$ or faster, it asymptotically dominates the true signals, and this implies underestimation. Next, suppose that $\mathfrak{K}$ grows slower than the signal energy to avoid asymptotically suppressing true signals and prevent underestimation. Since $V_i^*$ grows linearly with $N_s$, it asymptotically dominates any sublinear penalty $\mathfrak{K}$, this implies overestimation. In all scenarios, the error probability approaches unity.

In the case of an unknown noise level, $\Delta \tilde{L}_{\sigma i}^*$ is defined as $\Delta \tilde{L}_{\sigma i}^* = -\frac{N_s}{2}\ln(1 - l_{\sigma i}^{*2})$. 
If the condition $\bm{\omega}_{i}=\bm{\omega}_{0i }$ is not fulfilled, then, as follows from equations \eqref{le} and \eqref{depen}, the expected values $\mathrm{E}(l_{ci}^*)$ and $\mathrm{E}(l_{si}^*)$ depend on the true values of the signal amplitudes. Consequently, the numerator and the denominator of the ratio $l_{\sigma i}^{*2}$ depend on the true values of the signal amplitudes. Therefore, as the SNR or the number of samples tends to infinity, both the numerator and the denominator of $l_{\sigma i}^{*2}$ grow linearly with the SNR and the number of samples.
This linear growth cancels out in the ratio, thus the random variable $l_{\sigma i}^{*2}$ converges in probability to a constant $c_i \in (0, 1)$, which depends on the vectors $\mathbf{k}_{ci}$ and $\mathbf{k}_{si}$ defined in \eqref{depen}. As a result, as the SNR tends to infinity, $\Delta \tilde{L}_{\sigma i}^*$ converges to a positive constant $\Delta L_{\infty, i} = -\frac{N_s}{2}\ln(1 - c_i) > 0$. A data-independent penalty $\mathfrak{K}\nu$ cannot uniformly bound this constant for all possible amplitude values. It implies overestimation for sufficiently strong signals. Finally, as the number of samples tends to infinity, $\Delta \tilde{L}_{\sigma i}^*$ grows linearly with $N_s$. As in the known noise level case, a data-independent penalty $\mathfrak{K}\nu$ cannot simultaneously grow fast enough to suppress the linearly growing term $\Delta \tilde{L}_{\sigma i}^*$ and slow enough to preserve the true signals for all possible amplitude values. Therefore, as $N_s \to \infty$, the algorithm yields an incorrect model order. Thus, the error probability approaches unity.
\end{IEEEproof}

Suppose we design the MOS algorithm in the case of {\it a priori} known frequencies. If the error probability of this algorithm saves its behavior (as the SNR or number of samples tends to infinity) in the case of some errors in using frequencies values (i.e., the error probability, for example, still monotonically tends to zero or some non-zero constant), then we call such MOS algorithm the BL-robust algorithm.

Theorem \ref{negnegG} shows that non-adaptive MOS algorithms are not BL-robust. Note that, the numerical analysis in Section \ref{sims} shows that algorithms EEF, gMDL, NMDL, and RRT are also not BL-robust (see Fig. \ref{fig2} for the known noise level and Fig. \ref{fig2s} for the unknown noise level).

\begin{theorem} \label{PMEP_rob}
The PMEP-IR \eqref{RIT} and PMEP-I \eqref{IT} MOS algorithms are BL-robust.
\end{theorem}
\begin{IEEEproof}
Under known noise level and $\bm{\omega}_{i}^* \ne \bm{\omega}_{0i}$, \eqref{depen} implies $V_i^*$ grows linearly with $\lambda \in \{z_b^2, N_s\} \implies \lambda^{-1}V_i^* \xrightarrow{\Pr} c_i^* > 0$. Overestimation ($\hat{\nu} > \nu_0$) in \eqref{RIT} implies $V_{\nu_0+1}^* > \kappa_{\scriptscriptstyle IR} \max_{j} V_j^*$. Multiplying by $\lambda^{-1}$ as $\lambda \to \infty$ yields $c_{\nu_0+1}^* \ge \kappa_{\scriptscriptstyle IR} \max_j c_j^*$. Setting $\kappa_{\scriptscriptstyle IR} > c_{\nu_0+1}^* / \max_j c_j^*$ prevents overestimation.
Under unknown noise level, as $z_b \to \infty$, $\Delta \tilde{L}_{\sigma i}^* \xrightarrow{\Pr} \Delta L_{\infty, i} > 0$. Overestimation ($\hat{\nu} > \nu_0$) in \eqref{RIT} asymptotically yields $\Delta L_{\infty, \nu_0+1} \ge \kappa_{\scriptscriptstyle IR} \max_j \Delta L_{\infty, j}$. As $N_s \to \infty$, $N_s^{-1}\Delta \tilde{L}_{\sigma i}^* \xrightarrow{\Pr} C_i > 0$. Multiplying the overestimation inequality by $N_s^{-1}$ as $N_s \to \infty$ yields $C_{\nu_0+1} \ge \kappa_{\scriptscriptstyle IR} \max_j C_j$. Thus, bounding $\kappa_{\scriptscriptstyle IR}$ provides BL-robustness. The same reasoning applies to algorithm \eqref{IT}
\end{IEEEproof}

\subsection{Quasilikelihood performance analysis} \label{QLperfan}
Note that the performance analysis based on the formula \eqref{CDF_Slag} considers all statistical information about the ML estimates of the frequencies, and vice versa the analysis based on the formula \eqref{CDF1} does not take into account this information.
On the other hand, comparing results \eqref{CDF1}, \eqref{CDF_Slag} we can conclude that the theoretical performance analysis in the case of the ML estimation of the frequencies is quite more difficult than in the case of the BL estimation. Moreover, in practice, one may use some fast or simplified modifications of the maximum likelihood method for estimating the frequencies, thus the actual probability distribution of the estimates of frequencies can be very different from the distribution of the exact ML estimates. These reasons motivate us to propose a new approach to the performance analysis of the MOS algorithms, and this approach is called QL performance analysis.

Substitute $F_{i}^*\left( x \right)$ from \eqref{CDF1} to the formulas for the abridged error probabilities \eqref{AEP_GIC} \eqref{p_a_RIT2}, \eqref{p_a_I}. Note that these formulas are still valid for any values of $\bm{\omega}_i^*$ due to { Theorem~\ref{fieldOrh}}. Next, we can represent these probabilities as functions of the frequencies errors $p_{a{\scriptscriptstyle GIC}}\!\left(\Delta \bm{\omega} \right)$, $p_{a{\scriptscriptstyle IR}}\!\left(\Delta \bm{\omega} \right)$, $p_{a{\scriptscriptstyle I}}\!\left(\Delta \bm{\omega} \right)$,
where $\Delta \bm{\omega}=\bm{\omega}^*-\bm{\omega}_0$. Finally, we provide a performance measure that uses these functions only: 
$p_{aq}=\sum_{i=1}^{N_\Delta}p_a(\Delta \bm{\omega}_i)\mathfrak{L}(\Delta\bm{\omega}_i)$,
where $\lbrace\Delta \bm{\omega}_i \rbrace_{i=1}^{N_\Delta}$ is a set of the frequencies errors and $\mathfrak{L}(\Delta\bm{\omega})$ is a loss function (for instance, $\mathfrak{L}(\Delta\bm{\omega})=\Delta\bm{\omega}^2$).
The measure $p_{aq}$ shows the impact of the different errors in the used frequencies values on the MOS algorithm performance. The loss function allows specifying the errors contributions. Moreover, if  the probabilities of frequencies errors $p_{\Delta \bm{\omega}_i}$ are known, then one can also use them and calculate weighted average abridged error probability as $p_{waa}=\sum_{i=1}^{N_\Delta}p_{\Delta \bm{\omega}_i}p_a(\Delta \bm{\omega}_i)\mathfrak{L}(\Delta\bm{\omega}_i)$.

\subsection{Practical applications} \label{QLapp}
Let us provide some practical examples of the application of the QL approach. Consider situations when the frequencies of sinusoids in \eqref{OSModel} are {\it a priori} known, but {\it true} values $\omega_{0i}$ of these frequencies can deviate from the known values due to some causes. Describe some common causes for such deviations. The first cause is the Doppler effect. If the source (or sources) of the sinusoids in \eqref{OSModel} has some radial velocity relative to the receiver, then the frequencies $\omega_{0i}$ of the observed sinusoids in \eqref{OSModel} will be shifted from their known values \cite{doppapp1, doppapp2}.  The second cause is the hardware imperfection, for example, drift and/or offsets of frequencies of local oscillators \cite{oscapp1}.

In practice, the first way is to treat this case as a MOS problem with unknown frequencies, thus it is necessary to estimate the unknown frequencies and use these estimates in \eqref{MLFR} or \eqref{LFrSig}. This way involves computational difficulties. The second way is to use {\it a priori} known frequency values even though the true frequencies can deviate from them. Usually, the upper bound of possible deviations of the true frequency values from {\it a priori} known frequency values can be found. For example, it is usually possible to set some upper bound on the possible speed of the signal source in a given practical scenario. Using this upper bound of possible deviations and QL performance analysis, one can choose between the first and the second way (see, for example, Fig. \ref{fig4}). Note that if the second way is chosen, then one should use BL-robust MOS algorithms to avoid the problems described in Section \ref{BLROB}.

\section{Amplitudes unevenness} \label{PMEPfuture}
Formulas \eqref{strong_ce} indirectly point to one of the main problems of algorithms \eqref{RIT}, \eqref{IT} and other similar algorithms: their performance degrades with increasing the value of $\mathfrak{u}=\max a_{0i}/\min a_{0i}$. Let us call this ratio {\it amplitudes unevenness}. Using \eqref{Main_l_details} one can conclude that the increasing $\mathfrak{u}$ implies decreasing $\rho$. We have done a deeper numerical analysis of algorithms \eqref{RIT}, \eqref{IT} and it also approved the problem (for example, see Fig. \ref{fig5}). On the other hand, for example, GIC algorithms are free from such problems.  Let us propose a qualitative explanation of this phenomenon. First of all, we represent the decision function (see \eqref{AlModel}) as  
\begin{equation*}
\begin{split}
R&\left( \nu, x(t) \right) \!= \! \sum_{i=1}^{\nu}{ \left[ R\left( i, x(t) \right) \!-\! R\left( i \! - \! 1, x(t)\right) \right] }\\
&=\sum_{i=1}^{\nu}{ \left[\left(\tilde{\mathit{L}}_{i-1}-\tilde{\mathit{L}}_{i} \right)+(\mathfrak{P}_i-\mathfrak{P}_{i-1})\right]}= \sum_{i=1}^{\nu}{ \left[ -\frac{V_i}{2} + \mathfrak{p}_i \right]},
\end{split}
\end{equation*}
here we use representations \eqref{rep_add} and \eqref{Main_Rep}.
Let us characterize the contribution of individual terms $V_i$ and $\mathfrak{p}_i$ to the final value of decision function $R$ by the values of their expectations: $d_i^2=\mathrm{E}(V_{i})$ and $\mathfrak{d}_i=\mathrm{E}(\mathfrak{p}_i)$.

Next, we consider two types of penalty functions: local and global penalty functions. We assume that each term $\mathfrak{p}_i$ of the {\it local} penalty function can depend on the observed data $x(t)$ only through the term $V_{i}$. On the other hand, each term $\mathfrak{p}_i$ of the {\it global} penalty function can, for example, statistically depend on all terms in $\left\lbrace {V_j}\right\rbrace _{j=1}^{i}$ or in $\left\lbrace {V_j}\right\rbrace _{j=1}^{\nu_{max}}$. It implies that all values in the set $\left\lbrace {d_j^2}\right\rbrace _{j=1}^{i}$ or  $\left\lbrace {d_j^2}\right\rbrace _{j=1}^{\nu_{max}}$ can make some contribution to $\mathfrak{d}_i$.
This leads to both the advantages of the global penalty over the local one and the disadvantages. On the one hand, the global penalty can improve the performance of the MOS algorithm since the global penalty can enhance the growth of the decision function for $\nu>\nu_0$ (with SNR increasing). For the local penalty, if $\nu>\nu_0$, then $d_\nu^2=0$ due to Theorem~\ref{zeroE}. It is important to note that {\it one can extend Theorem \ref{negG} to the case of the local penalty term}.  On the other hand, the global penalty can make the MOS algorithm sensitive to the unevenness of the signal amplitudes, because the global penalty can suppress the contribution of the weak signals to the decision function if $\mathfrak{u} \gg 1$ (i.e., $\rho \ll 1$). For the local penalty, the terms corresponding to the strong signals do not affect the terms corresponding to the weak signals.

\begin{theorem} \label{robgen}
Let $\nu_0 \ge 2$.
1) If amplitude estimates are constrained to a domain bounded away from zero (e.g., $|\hat{a}_i| \ge \varepsilon > 0$), the algorithm loses SNR-consistency if there is at least one true amplitude $a_{0i}$, such that $0 < |a_{0i}| < \varepsilon/2$.
2) If amplitudes are unconstrained (i.e., can be equal to zero), then for any SNR-consistent MOS algorithm, if $\min_i a_{0i}^2 = \mathcal{O}(1)$ while $\max_i a_{0i}^2 \to \infty$, $p_e \to 1$.
3) Under unconstrained conditions, if the algorithm is BL-robust, $p_e \to 1$ as $z_b \to \infty$ and amplitude unevenness $\mathfrak{u}$ exceeds some finite threshold.
\end{theorem}

\begin{IEEEproof}
Suppose $\Delta R(\nu) = R(\nu) - R(\nu-1) = -\Delta L_\nu + \mathfrak{p}_\nu$ and $\mathfrak{p}_\nu \ge 0$. Let $\lambda_{max} \propto \max_i a_{0i}^2$ and $\lambda_{min} \propto \min_i a_{0i}^2$.

1) Assume $|\hat{a}_i| \ge \varepsilon > 0$. Consider $0 < |a_{0\nu_0}| < \varepsilon/2$. As $z_b \to \infty$, enforcing $|\hat{a}_{\nu_0}| \ge \varepsilon$ $\implies \Delta L_{\nu_0} \xrightarrow{\Pr} c < 0$. Since $\mathfrak{p}_{\nu_0} \ge 0$, $\Delta L_{\nu_0} < \mathfrak{p}_{\nu_0}$ holds asymptotically $\implies \Pr(\hat{\nu} < \nu_0) \to 1$.

2) Assume unconstrained amplitudes ($\Delta L_\nu = V_\nu/2$). SNR-consistency implies $\Pr(\hat{\nu} > \nu_0) \to 0$ as $\lambda_{max} \to \infty \implies \Pr(V_{\nu_0+1}/2 > \mathfrak{p}_{\nu_0+1}) \to 0$. By Theorem \ref{fieldOrh}, $V_{\nu_0+1} \sim \chi^2(2)$.  $\mathfrak{p}_{\nu_0+1} = \mathcal{O}_{\Pr}(1)$ yields $\Pr(V_{\nu_0+1}/2 > \mathfrak{p}_{\nu_0+1}) > 0$. Thus, to provide consistency we need $\mathfrak{p}_\nu \xrightarrow{\Pr} \infty$ as $\lambda_{max} \to \infty$. Preventing underestimation implies $\Pr(V_{\nu_0}/2 < \mathfrak{p}_{\nu_0}) \to 0$. If $\lambda_{min} = \mathcal{O}(1)$ as $\lambda_{max} \to \infty$, then $V_{\nu_0} = \mathcal{O}_{\Pr}(1)$. $\mathfrak{p}_{\nu_0} \xrightarrow{\Pr} \infty$ asymptotically dominates $V_{\nu_0} \implies \Pr(V_{\nu_0}/2 < \mathfrak{p}_{\nu_0}) \to 1 \implies p_e \to 1$.

3) Assume BL-robustness. By \eqref{depen}, $V_{\nu_0+1}$ grows linearly with $\lambda_{max} \implies \lambda_{max}^{-1}V_{\nu_0+1} \xrightarrow{\Pr} c_l > 0$. Preventing overestimation requires $\lambda_{max}^{-1}\mathfrak{p}_{\nu_0+1} \xrightarrow{\Pr} c_p \ge c_l/2 > 0$. Assuming linear global penalty scaling, $\lambda_{max}^{-1}\mathfrak{p}_{\nu_0} \xrightarrow{\Pr} c_p' > 0$. Preventing underestimation implies $V_{\nu_0}/2 > \mathfrak{p}_{\nu_0}$. Multiplying by $\lambda_{max}^{-1}$ as $\lambda_{max} \to \infty$ yields $(2\lambda_{max})^{-1}V_{\nu_0} \ge \lambda_{max}^{-1}\mathfrak{p}_{\nu_0} \xrightarrow{\Pr} c_p' > 0$. By \eqref{Main_l_details}, $\lambda_{min}^{-1}V_{\nu_0} \xrightarrow{\Pr} c_v > 0 \implies \lambda_{max}^{-1}V_{\nu_0} \xrightarrow{\Pr} c_v (\lambda_{min}/\lambda_{max}) = c_v \mathfrak{u}^{-2}$. Satisfying $c_v \mathfrak{u}^{-2}/2 \ge c_p'$ requires $\mathfrak{u} \le \sqrt{c_v / (2c_p')} = C$. If $\mathfrak{u} > C$, then $(2\lambda_{max})^{-1}V_{\nu_0} < \lambda_{max}^{-1}\mathfrak{p}_{\nu_0}$ asymptotically $\implies \Pr(\hat{\nu} < \nu_0) \to 1 \implies p_e \to 1$.
\end{IEEEproof}

\subsection{Complex penalty}
These observations indicate the advisability of using the local and global penalty functions together in a single complex penalty function, moreover, the contribution of the global part of the penalty function should be significant only for sufficiently large SNR (compared to the value of $1/\rho$).

Based on the results given in this paper, we introduce a weak optimality criterion (WOC) for MOS algorithms (instead of the minimum error probability criterion). According to WOC, the MOS algorithm must satisfy the following conditions.

WOC1:~Consistency (see \eqref{cons} for definition) at least for a certain range of tuning parameters.

WOC2:~BL-robustness (see Section \ref{BLrob}).

WOC3:~Assumption \eqref{assum_1}. 

WOC4:~Tuning parameters should be chosen to minimize the abridged error probability.

WOC5:~Low sensitivity to the amplitudes unevenness.

Next, let us describe the steps that need to be taken to provide optimality of algorithm \eqref{complexAlg} in the sense of WOC. Firstly, we need to find ranges of tuning parameters such that PMEP-C MOS algorithm \eqref{complexAlg} is consistent (see WOC1). One can directly use the methods presented in Section \ref{consistency} to find these ranges. Secondly, we need to find the values of the tuning parameters for which algorithm \eqref{complexAlg} satisfies BL-robustness (WOC2), WOC3, and low sensitivity to the amplitudes unevenness (WOC5) conditions taking into account the results of the first step. Essentially, it is necessary to choose (according to SNR) such a balance between the local and global penalties using tuning parameters so that algorithm \eqref{complexAlg} satisfies all these three conditions. Finally, we find values of tuning parameters that minimize the abridged error probability of algorithm \eqref{complexAlg} and belong to the ranges obtained as a result of the first two steps. One can use methods described in Section \ref{aep_calc} to find an analytical expression for the abridged error probability of algorithm \eqref{complexAlg}. This expression can help to solve problems formulated both in the second and final steps.

\begin{theorem} \label{Complex_Alg_Properties}
For the PMEP-C algorithm \eqref{complexAlg}:

1) Under the ML design approach, setting $\kappa_{1c} > 0$, $\kappa_{2c} > 0$, and $\kappa_{3c} \in (0, 1)$ provide consistency and robustness to amplitude unevenness (satisfying WOC5 for any $\rho > 0$).

2) Under the QL design approach, setting $\kappa_{3c} = 1$ provide BL-robustness (satisfying WOC2). Using $\kappa_{1c} > 0$ relaxes the lower bound on $\rho$ required for consistency compared to the PMEP-IR algorithm.
\end{theorem}
\begin{IEEEproof}
Let $\lambda \in \{N_s, z_b^2\}$ and $M = \max_i V_i$. As $\lambda \to \infty$, $\lambda^{-1}M \xrightarrow{\Pr} c_m > 0$ and $\lambda^{-1}V_{\nu_0} \xrightarrow{\Pr} c_{\nu_0} > 0$, with $c_{\nu_0}/c_m \ge \rho$. Consistency requires $V_{\nu_0+1} < g(M) < V_{\nu_0}$, where $g(M) = 2\kappa_{1c} + 2^{1-\kappa_{3c}}\kappa_{2c} M^{\kappa_{3c}}$.

1) Under ML, $\lambda^{-\gamma}V_{\nu_0+1} \xrightarrow{\Pr} 0 \,\, \forall \gamma > 0$. For $\kappa_{3c} \in (0, 1)$, $\lambda^{-1}g(M) \xrightarrow{\Pr} 0$. Multiplying $g(M) < V_{\nu_0}$ by $\lambda^{-1}$ as $\lambda \to \infty$ yields $0 \le c_{\nu_0}$, holding asymptotically $\forall \rho > 0$. Multiplying $V_{\nu_0+1} < g(M)$ by $\lambda^{-\kappa_{3c}}$ as $\lambda \to \infty$ yields $0 \le 2^{1-\kappa_{3c}}\kappa_{2c} c_m^{\kappa_{3c}}$, holding $\forall \kappa_{2c} > 0$.

2) Under QL, $\lambda^{-1}V_{\nu_0+1} \xrightarrow{\Pr} c_{\epsilon} > 0$. If $\kappa_{3c} < 1$, multiplying $V_{\nu_0+1} < g(M)$ by $\lambda^{-1}$ as $\lambda \to \infty$ yields $c_{\epsilon} \le 0$ (overestimation). Thus, BL-robustness requires $\kappa_{3c} = 1 \implies g(M) = 2\kappa_{1c} + \kappa_{2c} M$. Multiplying $V_{\nu_0+1} < g(M) < V_{\nu_0}$ by $\lambda^{-1}$ as $\lambda \to \infty$ yields $c_{\epsilon} \le \kappa_{2c} c_m \le c_{\nu_0} \implies c_{\epsilon}/c_m \le \kappa_{2c} \le c_{\nu_0}/c_m$. Since $c_{\nu_0}/c_m \ge \rho$, this holds if $c_{\epsilon}/c_m < \kappa_{2c} < \rho$. Parameter $\kappa_{1c} > 0$  allows $\kappa_{2c} \to c_{\epsilon}/c_m$, relaxing the bound $\kappa_{\scriptscriptstyle IR} < \rho$ from \eqref{strong_ce}.
\end{IEEEproof}

\subsection{Adaptation of tuning parameters} \label{adtp}
Note that the described problem of the high sensitivity of  PMEP MOS algorithms \eqref{RIT} and \eqref{IT} to amplitudes unevenness can be solved in a different way for a wide range of practical applications. 
Suppose that $z_1, z_2, ...$ are true values of SNRs and they can tend to infinity independently (not only as specified in Section \ref{consistency}).  Let the parameter $\mathfrak{u}$ be known or can be estimated or can be bounded in some way. In 
this case, one can fix the high sensitivity of  PMEP MOS algorithms \eqref{RIT} and \eqref{IT} to amplitudes unevenness using the true value of $\mathfrak{u}$ (or its estimate, or its lower bound). For example, in the case of PMEP-IR algorithm \eqref{RIT}, we should change the tuning parameter $\kappa_{IR}$ to the following $\kappa_{IR}/{\breve{\mathfrak{u}}}$, where $\breve{\mathfrak{u}}$ is a true value or an estimate or a lower bound of the parameter $\mathfrak{u}$. 

\section{Simulations} \label{sims}
In this section, we numerically study the problem of estimating the number of superimposed sinusoids as a special case of problem \eqref{OSModel}, thus we assume that $f_i(t)=1$ and ${\Psi _i}(t)=0$ for all $i$. We use settings and notations similar to those of \cite{sim, Djuric1996, Nadler2011, Mariani2015}. We suppose that $\omega_{0i}=2\pi(0.2+(i-1)/{N_s})$, $N_s = 128$, $\varphi_{01}=0$, $\varphi_{02}=-\pi/8$, $\varphi_{03}=-\pi/6$, for $i>3$ $\varphi_{0i}=-\pi/(i+5)$, and $\forall i$  $a_{0i}=a$.  
For all simulations, it is supposed (without loss of generality) that $\sigma_0=1$, thus we tune the SNRs \eqref{SNR_def} by the amplitudes only.
We provide simulations for six cases: {\it a priori} known frequencies, unknown frequencies (each iteration requires ML estimation of frequencies), partially unknown frequencies (in this case we use the BL estimates of frequencies only), and for {\it a priori} known and unknown noise level.
In the case of unknown frequencies, one has to numerically estimate these frequencies, however, it is a complex computational problem. We use an approach that is similar to that presented in \cite{sim} to resolve these difficulties.

Besides the above settings, we need to find optimal in the sense of minimum error probability values of the tuning parameters for the numerical studies of the PMEP algorithms.
Firstly, we find 
\begin{equation} \label{kappaopt}
\kappa_{\scriptscriptstyle IR}^{opt/abr}=\arg \mathop {\min }\limits_{\kappa_{\scriptscriptstyle IR}}\,{p_{a{\scriptscriptstyle IR}}\!\left( \kappa_{\scriptscriptstyle IR}\right) },\, \kappa_{\scriptscriptstyle I}^{opt/abr}=\arg \mathop {\min }\limits_{\kappa_{\scriptscriptstyle I}}\,{p_{a{\scriptscriptstyle I}}\!\left( \kappa_{\scriptscriptstyle I}\right) },
\end{equation}
where abridged error probabilities ${p_{a{\scriptscriptstyle IR}}\!\left( \kappa_{\scriptscriptstyle IR}\right) }$ and ${p_{a{\scriptscriptstyle I}}\!\left( \kappa_{\scriptscriptstyle I}\right) }$ are defined by formulas \eqref{p_a_RIT2} and \eqref{p_a_I}, respectively, in the case of $\mathcal{N}=5$, $\nu_{0}=3$, and $\mathrm{SNR}=-4 \, \mathrm{dB}$. Moreover, we find optimal values of tuning parameters $\kappa_{\scriptscriptstyle IR}$ and $\kappa_{\scriptscriptstyle I}$ in limited ranges obtained in \eqref{strong_ce} and \eqref{strong_ce2} to ensure that resulting MOS algorithms with tuning parameters \eqref{kappaopt} are consistent. 
Secondly, we use numerical methods to calculate error probabilities  ${p_{e{\scriptscriptstyle IR}}\!\left( \kappa_{\scriptscriptstyle IR}\right) }$ and ${p_{e{\scriptscriptstyle I}}\!\left( \kappa_{\scriptscriptstyle I}\right) }$  of algorithms \eqref{RIT} and \eqref{IT} in the case of unknown frequencies, $\mathcal{N}=5$, $\nu_{0}=3$, and $\mathrm{SNR}=-4 \, \mathrm{dB}$. Next, we find  
$\kappa_{\scriptscriptstyle IR}^{opt}=\arg \mathop {\min }\limits_{\kappa_{\scriptscriptstyle IR}}\,{p_{e{\scriptscriptstyle IR}}\!\left( \kappa_{\scriptscriptstyle IR}\right) },\, \kappa_{\scriptscriptstyle I}^{opt}=\arg \mathop {\min }\limits_{\kappa_{\scriptscriptstyle I}}\,{p_{e{\scriptscriptstyle I}}\!\left( \kappa_{\scriptscriptstyle I}\right) },$ 
Finally, we obtain the following result
$$\kappa_{\scriptscriptstyle IR}^{opt} \approx \kappa_{\scriptscriptstyle IR}^{opt/abr}\!=0.25,\,\,\, \kappa_{\scriptscriptstyle I}^{opt} \approx \kappa_{\scriptscriptstyle I}^{opt/abr}\!=3.$$ This result and our additional studies lead us to the following {\it conjecture: one can use relatively simple formulas \eqref{CDF1} with the formulas for the abridged error probability (for example, with \eqref{p_a_RIT2} and \eqref{p_a_I}) to find optimal values of the tuning parameters in the cases of known and unknown frequencies.}

We choose algorithm \cite[Eq. 20]{Nadler2011} as an example of a GIC algorithm for our numerical studies. The structure of this algorithm includes a tuning parameter $\alpha$. We use this algorithm with $\alpha=0.007$. Note that the RRT algorithm \cite[Eq. 2]{RRT} also contains a tuning parameter denoted as $\alpha$ in its structure. We use the RRT algorithm with $\alpha=0.01$.

\begin{figure}[!t]
	\centering
	\includegraphics[width=3.2in]{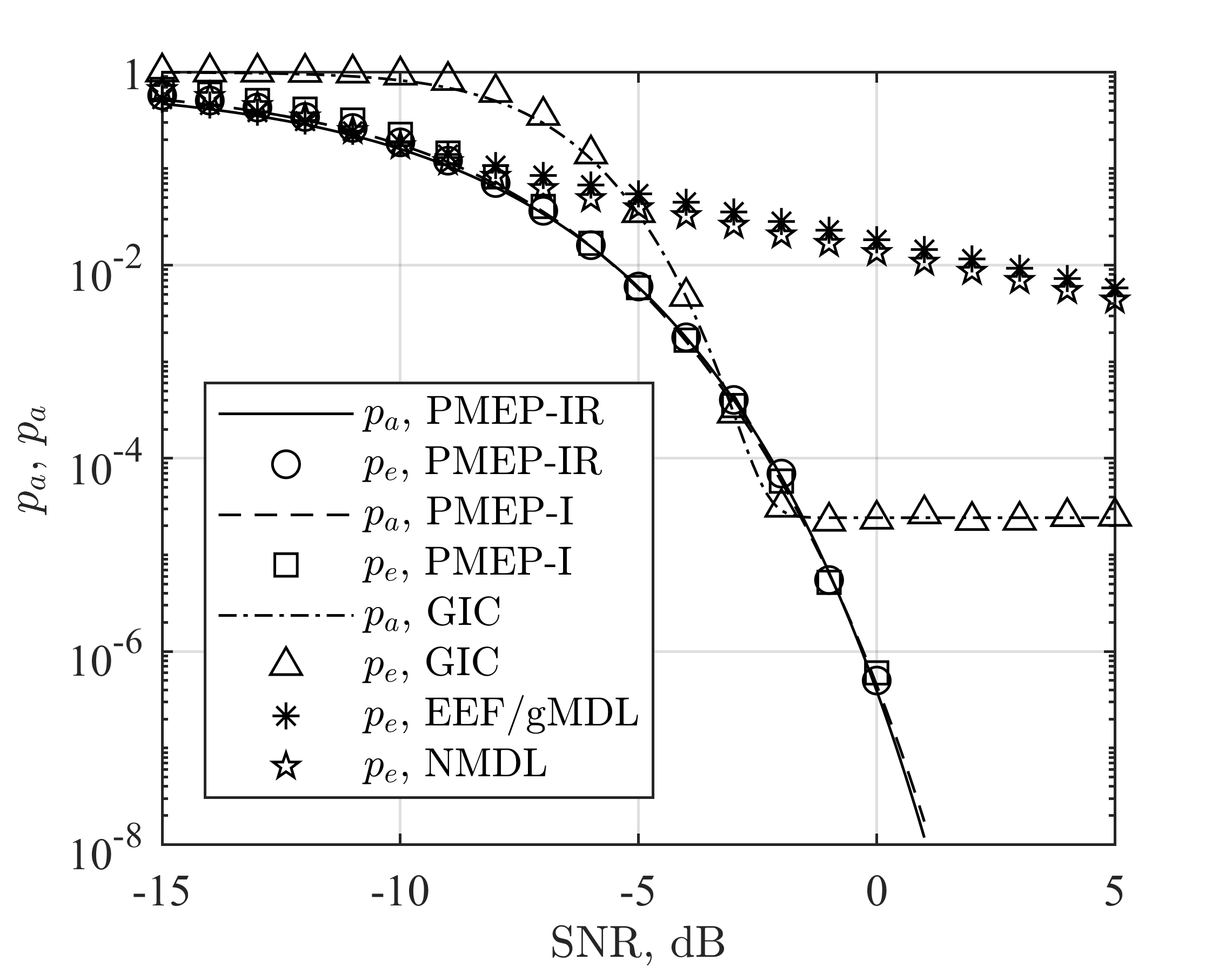}
	\caption{Abridged error probabilities $p_a$ and error probabilities $p_e$ of the different MOS algorithms versus SNR. {A priori} known frequencies and noise level case.}
	\label{fig1}
\end{figure}

\begin{figure}[!t]
	\centering
	\includegraphics[width=3.2in]{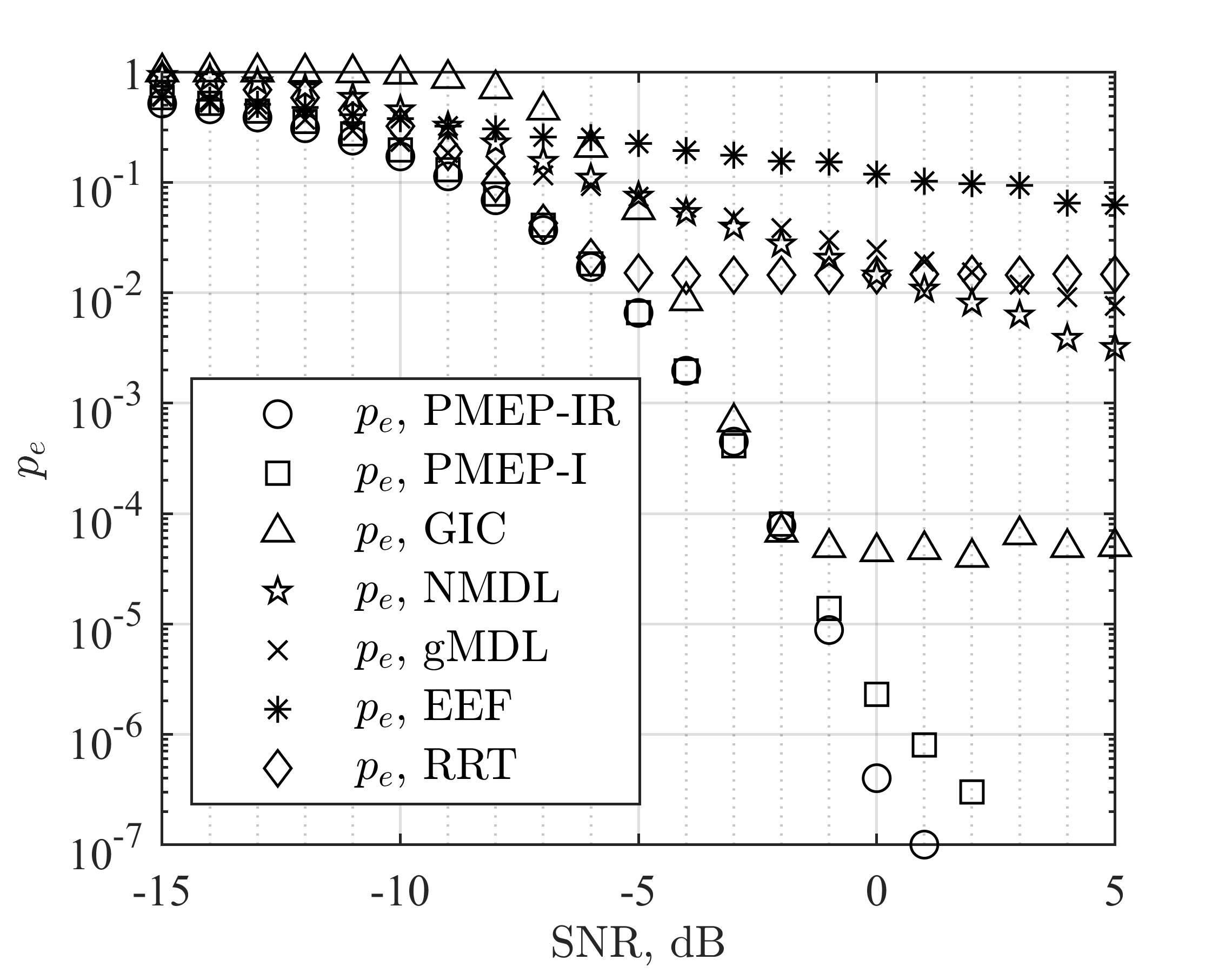}
	\caption{Error probabilities $p_e$ of the different MOS algorithms versus SNR. {A priori} known frequencies and unknown noise level case.}
	\label{fig1s}
\end{figure}

\begin{figure}[!t]
	\centering
	\includegraphics[width=3.2in]{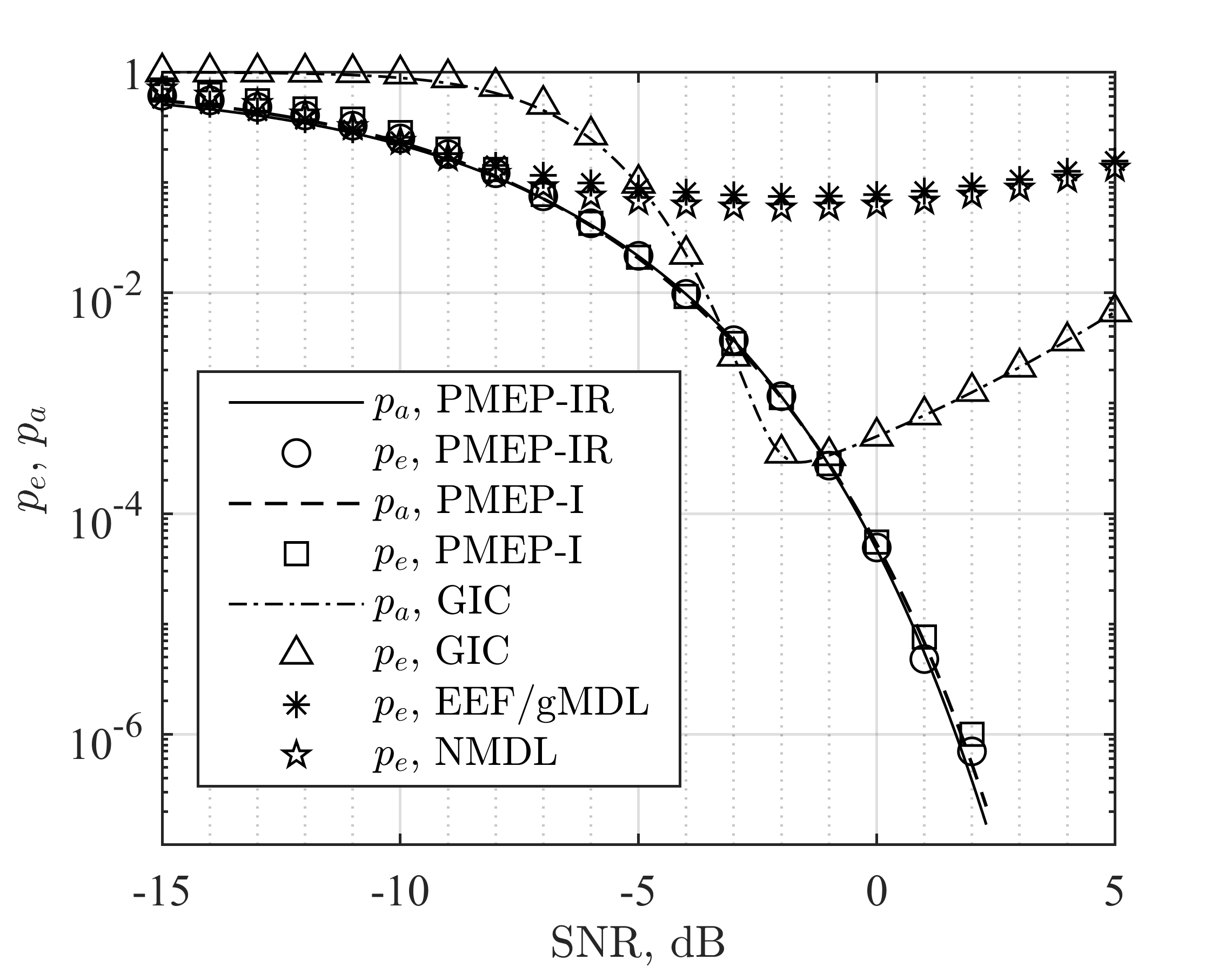}
	\caption{Abridged error probabilities $p_a$ and error probabilities $p_e$ of the different MOS algorithms versus SNR. In this figure, all presented MOS algorithms are designed within the QL approach. Here, we suppose that $\Delta \omega= 0.0025$. {A priori} known noise level case.}
	\label{fig2}
\end{figure}

\begin{figure}[!t]
	\centering
	\includegraphics[width=3.2in]{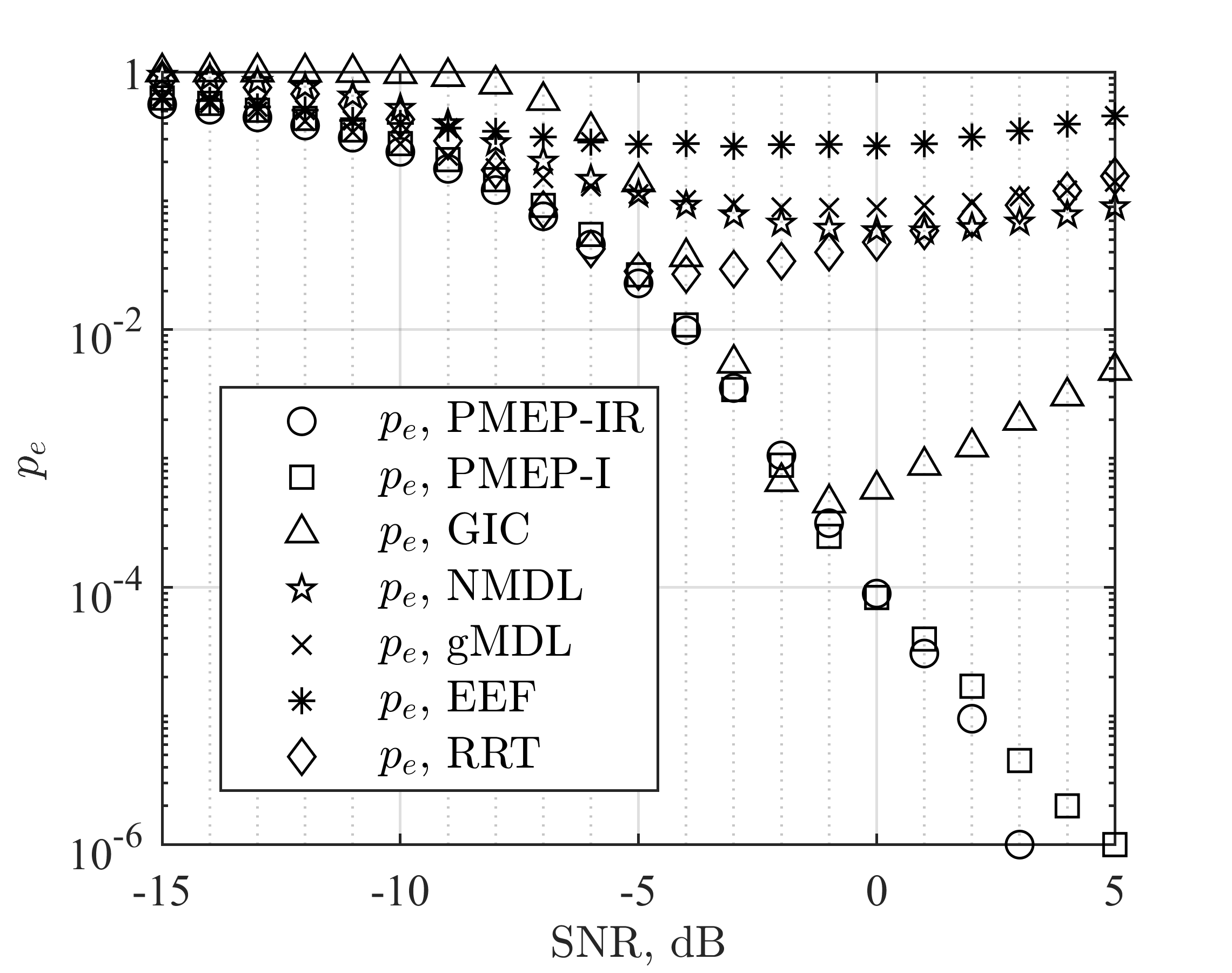}
	\caption{Error probabilities $p_e$ of the different MOS algorithms versus SNR. In this figure, all presented MOS algorithms are designed within the QL approach. Here, we suppose that $\Delta \omega= 0.0025$. {A priori} unknown noise level case.}
	\label{fig2s}
\end{figure}

\begin{figure}[!t]
	\centering
	\includegraphics[width=3.2in]{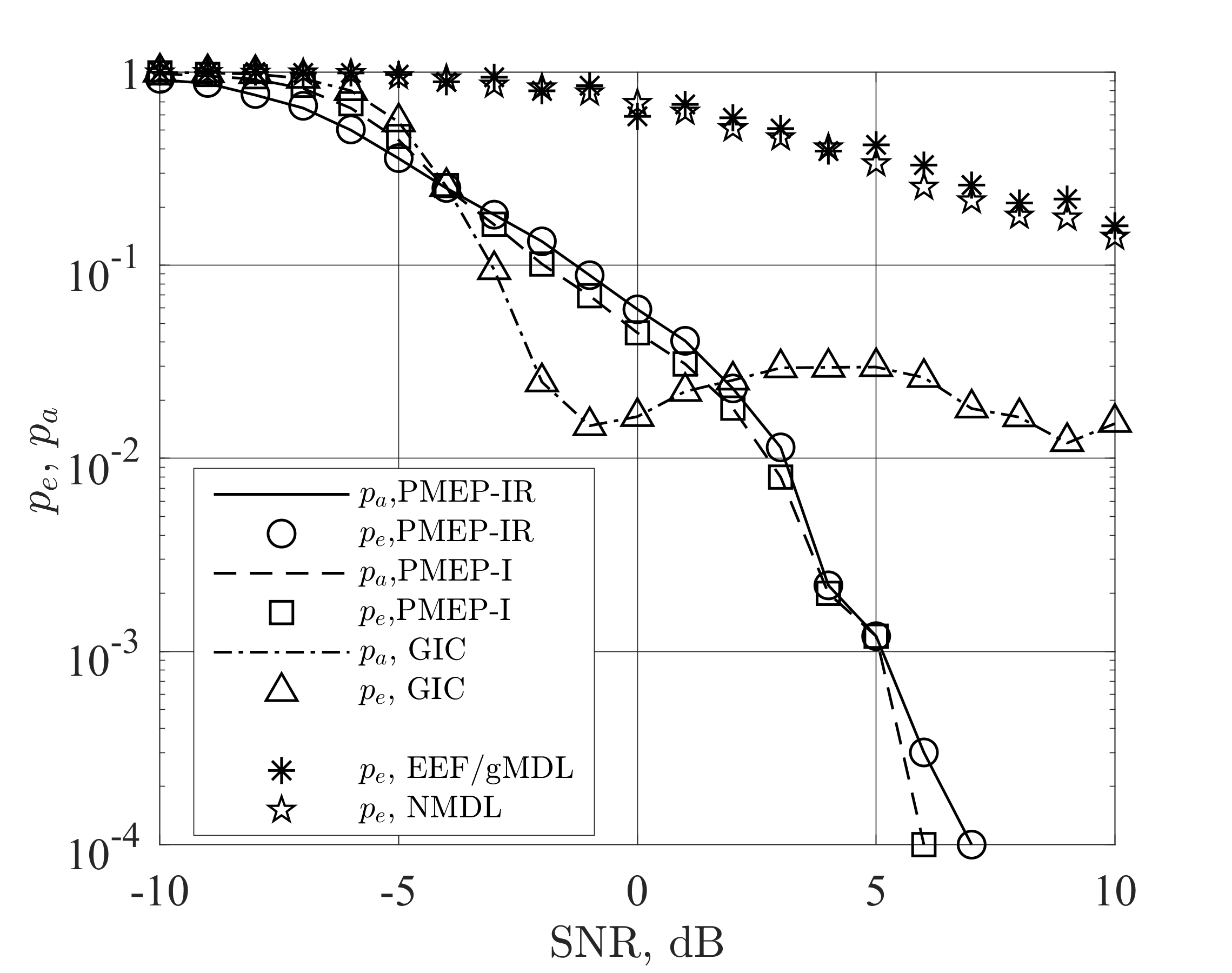}
	\caption{Abridged error probabilities $p_a$ and error probabilities $p_e$ of different MOS algorithms versus SNR. We use the ML approach to design all the MOS algorithms presented in this figure.}
	\label{fig3}
\end{figure}

\begin{figure}[!t]
	\centering
	\includegraphics[width=3.2in]{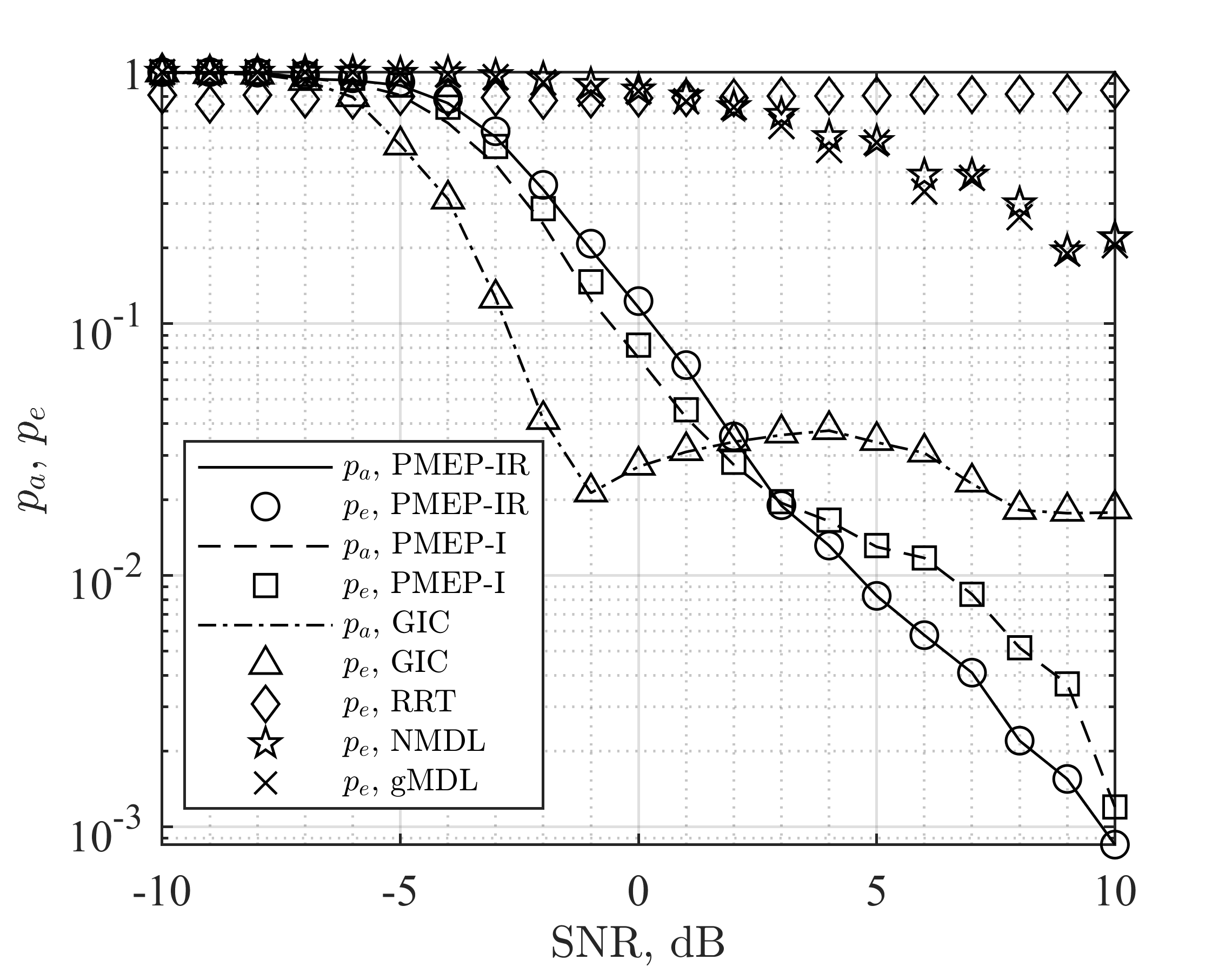}
	\caption{Abridged error probabilities $p_a$ and error probabilities $p_e$ of different MOS algorithms versus SNR. We use the ML approach to design all the MOS algorithms presented in this figure. {A priori} unknown noise level case.}
	\label{fig3s}
\end{figure}

\begin{figure}[!t]
	\centering
	\includegraphics[width=3.2in]{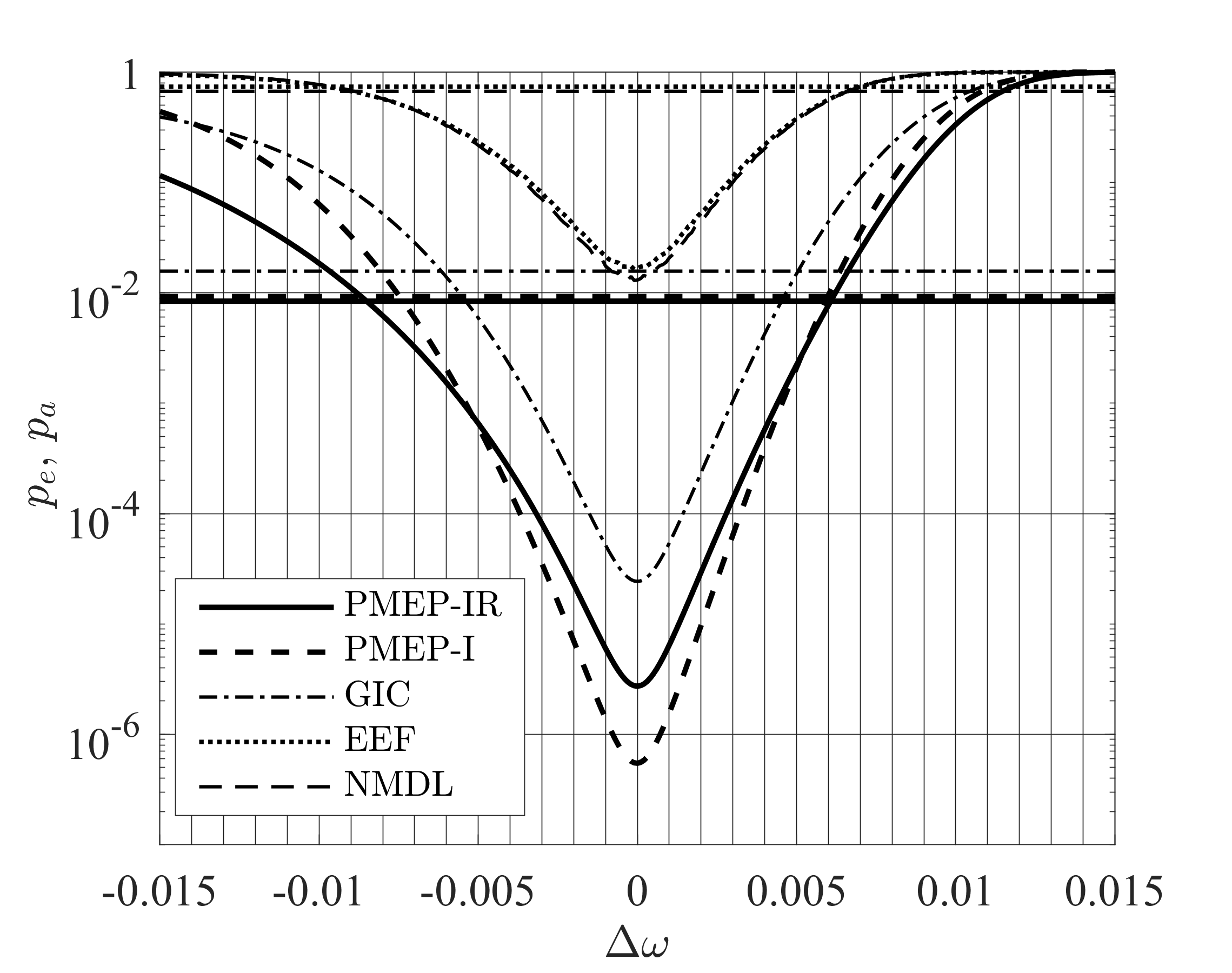}
	\caption{Error probabilities $p_e$ versus frequency error $\Delta \omega$. A comparison of the QL and ML (horizontal bold lines) approaches.}
	\label{fig4}
\end{figure}

\begin{figure}[!t]
	\centering
	\includegraphics[width=3.2in]{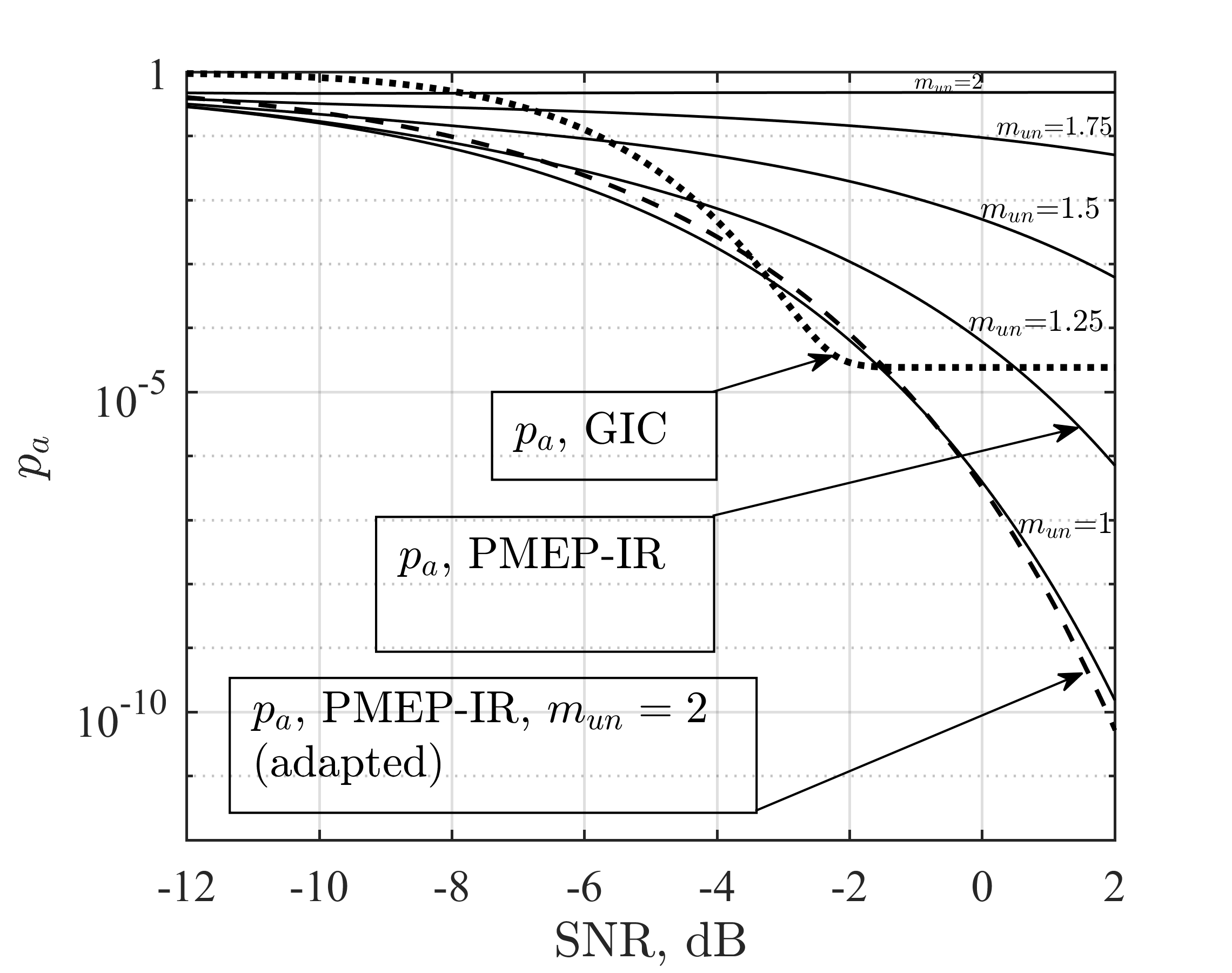}
	\caption{Abridged error probabilities $p_a$ of the different MOS algorithms versus SNR.}
	\label{fig5}
\end{figure}

We set $\mathcal{N}=5, \nu_{0}=3$ for numerical studies depicted in Fig. \ref{fig1} - \ref{fig5}.

Firstly, we study the case of {\it a priori} known frequencies and {\it a priori} known noise level. Figure \ref{fig1} shows abridged error probabilities $p_a$ theoretically calculated for the PMEP-IR (see \eqref{p_a_RIT2}), PMEP-I \eqref{p_a_I}, GIC \eqref{AEP_GIC} algorithms and error probabilities $p_e$ calculated numerically for PMEP-IR, PMEP-I, GIC, EEF/gMDL, NMDL algorithms, all as functions of SNR (g-MDL and EEF algorithms have the same performance in the case of {\it a priori} known noise level). Fig. \ref{fig1} confirms Theorem~\ref{negG} (see the dash-dotted line and triangles), i.e., it confirms that the GIC algorithm is not SNR-consistent. Also, Fig. \ref{fig1} shows that starting with $\textrm{SNR}=-6\textrm{ dB}$ the error probability can be well approximated using the abridged error probability. Despite the fact that MOS algorithm \eqref{GIC_R} does not satisfy condition \eqref{assum_1}, the abridged error probability is still an accurate approximation for the error probability of this algorithm (see the dash-dotted line and triangles). Next, Figure \ref{fig1s} shows error probabilities $p_e$ calculated numerically for PMEP-IR, PMEP-I, GIC, EEF, gMDL, NMDL algorithms in the case of {\it a priori} known frequencies and unknown noise level, all as functions of SNR. Note that, in the case of the unknown noise level we use the form of EEF algorithm from \cite{Kay2}.  


Now let us study the QL approach to the design of MOS algorithms (the case of partially unknown frequencies). Figures \ref{fig2} and \ref{fig2s} show the same dependencies with the same settings as in Fig. \ref{fig1} and Fig. \ref{fig1s}, respectively. However, in this case, we use the BL estimates
$\omega_i^*=\omega_{0i}+\Delta \omega$ (with the error in all BL estimates $\Delta \omega= 0.0025 \approx 5\%(\omega_{0i+1}-\omega_{0i})$) instead of $\omega_{0i}$ to design all the studied MOS algorithms. Fig. \ref{fig2} confirms (see the dash-dotted line and triangles) Theorem \ref{negnegG}, i.e., it confirms that the GIC algorithm is not robust to errors in used frequencies values. Moreover, this figure shows that the EEF, gMDL, and NMDL algorithms are also not robust to the errors in used frequencies values (see asterisks and pentagrams). Figure \ref{fig2s} confirms these conclusions in the case of the unknown noise level. Also, in this particular case, Fig. \ref{fig2} shows that starting with $\textrm{SNR}=-4\textrm{ dB}$ the error probability can also be well approximated by the abridged error probability. 

Figures \ref{fig3} and \ref{fig3s} show our results in the case of {\it a priori} unknown frequencies (the ML approach) and {\it a priori} known and unknown noise level, respectively. These figures show numerically calculated values of abridged error probabilities $p_a$ and error probabilities $p_e$ of the MOS algorithms versus SNR. The values of probabilities $p_a$ and $p_e$ are calculated based on the identical realizations of the observed data, thus the statistical variations of these values are the same. Figures \ref{fig3} and \ref{fig3s} confirm the abridged error probability as a good approximation of the error probability starting with $\textrm{SNR}=-6 \textrm{dB}$.

Note that, from our theoretical and numerical studies, partially shown in Figures \ref{fig2} - \ref{fig3s}, it follows that the MOS algorithms that demonstrate poor performance in the case of BL estimate (see Fig. \ref{fig2} and \ref{fig2s}) also demonstrate poor performance in the case of ML estimate of frequencies (see Fig. \ref{fig3} and \ref{fig3s}) and vice versa. This fact confirms the QL performance analysis proposed in Section \ref{QLperfan}.

Overall figures \ref{fig1} - \ref{fig3s} show a significant advantage of the studied PMEP algorithms (PMEP-IR, PMEP-I) in the case of $\mathrm{SNR} >0$  (design with {\it a priori} known frequencies and the QL design approach) and in the case of $\mathrm{SNR} >4$  (the ML design approach). This advantage can be crucial in many practical applications.

Next, compare the QL and ML approaches to the design of MOS algorithms in the case of the known noise level. Figure \ref{fig4} shows the abridged error probabilities of GIC, PMEP-I, and PMEP-IR algorithms and error probabilities of EEF and NMDL algorithms in the case of the QL approach versus errors in the used BL frequencies estimates $\Delta \omega_i$. In Fig. \ref{fig4} we suppose that for all $i$: $\Delta \omega_i=\Delta \omega$.  Also, in Fig. \ref{fig4} we plot the error probabilities of the same MOS algorithms, but in the case of the ML design approach (horizontal bold lines). Note that in Fig. \ref{fig4} the lines of the same type refer to the same algorithms. For all the cases we set $\mathrm{SNR}=0\mathrm{ dB}$. Solid and dashed horizontal lines almost coincide.

Comparing the plotted curves in Fig. \ref{fig4} in pairs (a pair of curves of the same type corresponds to the same MOS algorithm), we can conclude that for each MOS algorithm, there is an interval of errors $\Delta \omega_i$ such that the MOS algorithm designed using the QL approach has better performance than one designed using the ML approach. We call these intervals the BL intervals. If all possible errors in frequencies estimates in a particular practical scenario are less than the corresponding BL intervals, then one can use the BL estimates instead of the ML estimates to design the MOS algorithm without loss of its performance. Here we list the BL intervals obtained from Fig. \ref{fig4}: the BL interval is $0.006$ for the PMEP-IR algorithm; $0.006$ for the PMEP-I algorithm;  $0.005$ for the GIC algorithm; $0.007$ for the EEF, gMDL, NMDL algorithms. 
Comparing these values of BL intervals and the value of the error in the BL estimates $\Delta \omega= 0.0025$ which is chosen for numerical simulations presented in Fig. \ref{fig2} one concludes that the robustness problem described in Theorem \ref{negnegG} has a practical meaning. Next, consider these results according to practical applications from Section \ref{QLapp}. Suppose that, deviations of frequencies from their {\it a priori} known values occur due to the motion of the source (or sources) of observed signal \eqref{OSModel}. Using the Doppler shift formula (see for example \cite{doppapp1}) one can conclude that if all speeds of sources satisfy the following condition $v_{source}< 0.0039c$ (here $c$ -- is the propagation speed of waves in the medium), then one can use the QL approach instead of the ML approach without any losses in performance. Let us clarify this condition for the case of electromagnetic waves in a vacuum: $v_{source} < 10^6 \text{ m/s}$. Thus, in this case, one can use BL estimates instead of ML estimates for all practical applications.

Study the impact of amplitudes unevenness on the performance of algorithms PMEP-I and PMEP-IR. We introduce amplitudes unevenness to our model with a multiplier $m_{un}\ge 1$ as follows: $a_{01}=m_{un}a$, while $\forall i > 1$:  $a_{0i}=a$, i.e., $\mathfrak{u}=m_{un}$. Fig. \ref{fig5} shows abridged error probabilities $p_a$ calculated for PMEP-IR \eqref{p_a_RIT2} algorithm (performance of PMEP-I show a similar behavior) for different values of $m_{un} \in \left\lbrace 1, 1.25, 1.5, 1.75, 2 \right\rbrace$ and for GIC algorithm \eqref{AEP_GIC} for comparison.
Fig. \ref{fig5} confirms the high sensitivity of algorithms containing only the global penalty function to amplitudes unevenness (see a group of thin solid lines). On the other hand, the bold dashed line shows the performance of the PMEP-IR algorithm for $m_{un}=2$ and in the case of using the tuning parameter adaptation, as suggested in Section \ref{adtp}. Comparing the solid and dashed lines, we conclude that the use of an adapted tuning coefficient ${\breve{\mathfrak{u}}}\kappa_{IR}$ can remove the sensitivity of the PMEP-IR algorithm to amplitude unevenness. Our additional studies show that the amplitudes unevenness problem is still partially valid in the case of the unknown noise level. One can explain this using an analysis similar to the one given in Section \ref{PMEPfuture}. However, there are some differences between the case of known noise level and the case of the unknown noise level. For example, if in the situation shown in Fig. \ref{fig5} the noise level is unknown, then the PMEP-IR MOS algorithm will retain its performance for any value of amplitudes unevenness and without any modification.

\begin{figure}[!t]
	\centering
	\includegraphics[width=3.2in]{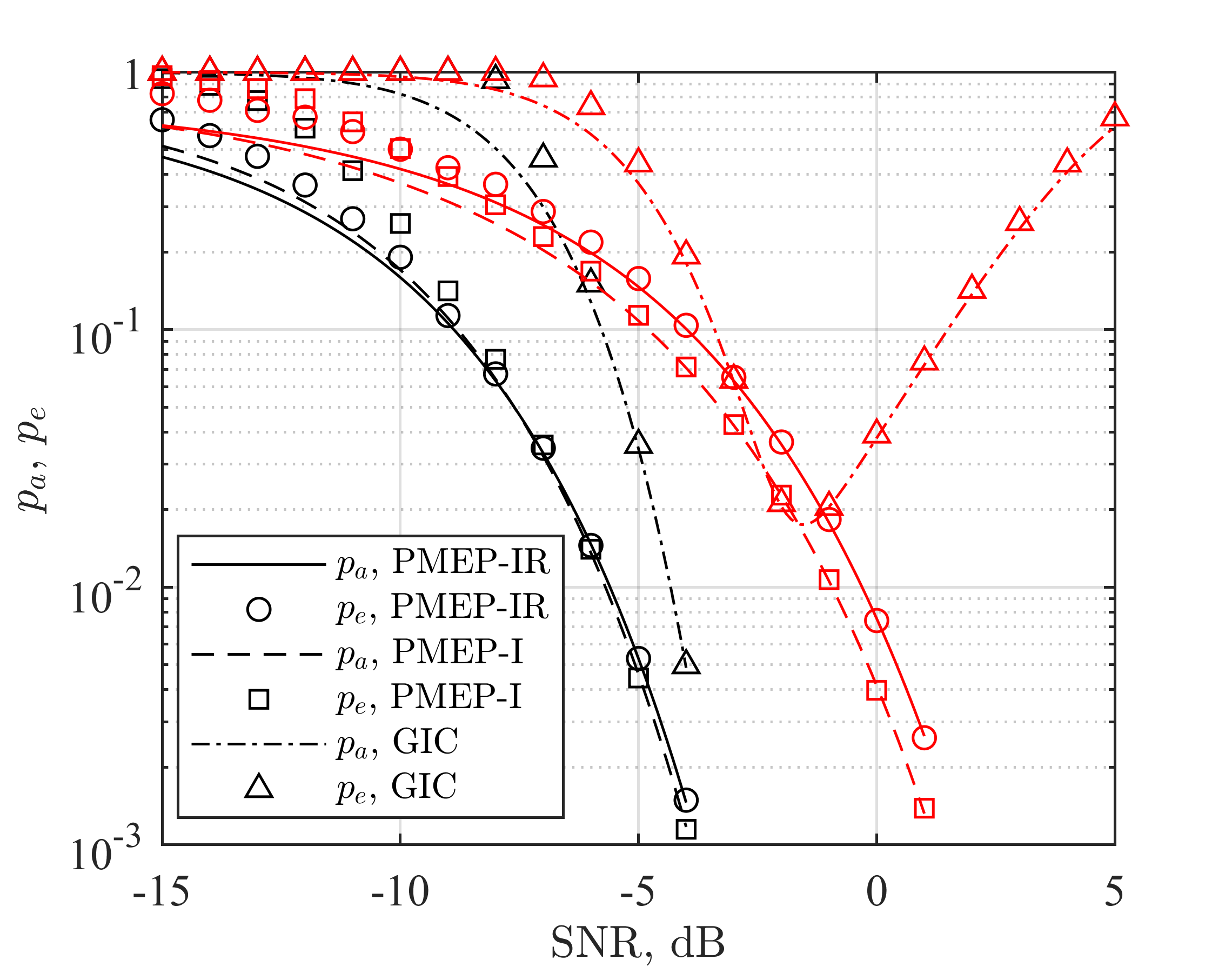}
	\caption{Abridged error probabilities $p_a$ and error probabilities $p_e$ of the different MOS algorithms versus SNR. {A priori} known frequencies case and partially unknown frequencies case (red color). For the partially unknown frequencies case, we suppose that $\Delta \omega= 0.0025$. {A priori} known noise level.}
	\label{fig9}
\end{figure}

Finally, let us study the quality of approximation of the error probability by the abridged error probability for big values of $\mathcal{N}$ and $\nu_{0}$. We study the cases of {\it a priori} known and partially unknown frequencies. We set $\mathcal{N}=50, \nu_{0}=25$ for numerical studies depicted in Fig. \ref{fig9}. Figure \ref{fig9} shows abridged error probabilities $p_a$ theoretically calculated for the PMEP-IR (see \eqref{p_a_RIT2}), PMEP-I \eqref{p_a_I}, GIC \eqref{AEP_GIC} algorithms and error probabilities $p_e$ calculated numerically for PMEP-IR, PMEP-I, and GIC algorithms, all as functions of SNR. For the case of partially unknown frequencies, we suppose that $\Delta \omega= 0.0025$ and we use red color to present results for this case. 
Fig. \ref{fig9} shows that  the error probability can be well approximated using the abridged error probability starting with $\textrm{SNR}=-6\textrm{ dB}$ in the case of {\it a priori} known frequencies and starting with $\textrm{SNR}=-4\textrm{ dB}$ in the case of partially unknown frequencies. Thus, the quality of the approximation is almost independent of values of $\mathcal{N}$ and $\nu_{0}$.

\section{Conclusions} \label{concl}
In this paper, we proposed new approaches to the design, consistency analysis, and performance analysis of MOS algorithms and applied these approaches to the classical problems of estimating the number of modulated sinusoids with unknown amplitudes, phases, known, unknown, and partially unknown frequencies in the presence of noise with known and unknown levels. 

\appendices

\section{Preliminary results} \label{prelim}
\subsection{A new non-iterative formula for the Gram-Schmidt process} \label{newNIGS}
Let $\mathbb{V}$ be a vector space over the field $\mathbb{C}$ with an inner product $ \left\langle {\cdot,\cdot} \right\rangle $. Suppose $A_{1}, \ldots ,A_{N}$ are linearly independent vectors in $\mathbb{V}$. Let $n$ and $N$ be natural numbers, and $1 \le n \le N$.  Suppose for any $n$, ${{\mathbf{G}}_n} = (G_{i,j})_{i,j=1}^n \in  \mathbb{C}^{ n \times n} $ is a Gram matrix with dimension $n \times n$ whose element  $ {G_{i,j}} = \left\langle {A_i},{A_j} \right\rangle $  is the inner product of $A_i$ and $A_j$, ${{\mathbf{g}}_n}=\left( {G_{1,n}, \ldots ,G_{n-1,n}} \right)^T \in  \mathbb{C}^{n-1 \times 1}$. 

Now let us define vectors $B_{1} \in \mathbb{V}, \ldots ,B_{N} \in \mathbb{V}$ as
\begin{equation} \label{MainFormula}
B_n =\left( A_n - \sum\limits_{i = 1}^{n-1} {A_i \left({\mathbf{G}_{n - 1}^{ - 1}}{\mathbf{g}_n} \right)_i}  \right)/\sqrt {S_n} ,
 \end{equation} 
if  $2 \le n \le N$, $B_1 = A_1/\sqrt {G_{1,1}}$.  

Here $S_n = {G_{n,n}} - {\mathbf{g}}_n^ H {\mathbf{G}}_{n - 1}^{ - 1}{{\mathbf{g}}_n}$ is the Schur complement \cite{MTG}, the superscript $H$ denotes the Hermitian transpose,  $({\mathbf{G}}_{n - 1}^{ - 1}{{\mathbf{g}}_n})_i$ denotes the $i$-th component of the vector ${\mathbf{G}}_{n - 1}^{ - 1}{{\mathbf{g}}_n}$.

Let us prove that formulas \eqref{MainFormula} define the Gram-Schmidt orthonormalization process. 
\begin{theorem} \label{NewfGS}
Suppose that $\tilde B_{1}, \ldots , \tilde B_{N}$ are orthonormal vectors produced by applying the Gram-Schmidt process to the vectors $A_{1}, \ldots ,A_{N}$, and that the vectors $B_1, \ldots, B_N$ are defined by equations \eqref{MainFormula}; then $\tilde B_n = B_n$ for any $1 \le n \le N$.
\end{theorem}
\begin{IEEEproof} 
Please see Appendix~\ref{Proof1}.
	\end{IEEEproof}
\subsection{Preliminary probabilistic results} \label{prelprob}

Let $\mathcal{X}_{1}, \ldots ,\mathcal{X}_{N}$ be linearly independent real valued random variables with finite variances.  Suppose for any $1 \le n \le N$, $\bm{\mathcal{X}}_n = \left( {\mathcal{X}_{1}, \ldots ,\mathcal{X}_{n}} \right)^T$ is a random vector, ${\bm{\mathcal{C}}_n} = (\mathcal{C}_{i,j})_{i,j=1}^n \in  \mathbb{R}^{ n \times n} $ is a covariance matrix of the vector  $\bm{\mathcal{X}}_n$ (the element  $ {\mathcal{C}_{i,j}} = {\mathop{\mathrm {cov}}} ({\mathcal{X}_i},{\mathcal{X}_j}) $ is the covariance between $\mathcal{X}_i$ and $\mathcal{X}_j$), ${{\bm{\mu}}_n}=\left( {\mu_1, \ldots ,\mu_n} \right)^T$ is the mean vector of $\bm{\mathcal{X}}_n$.

Despite the fact that random variables with arbitrary expectations do not form a vector space, one can find that applying the Gram-Schmidt process to random variables $\mathcal{X}_{1}, \ldots ,\mathcal{X}_{N}$ produces uncorrelated random variables with unit variances. Thus Theorem \ref{NewfGS} implies the following result. 
\begin{theorem} \label{randOrh}
If random variables $\{\mathcal{Y}_{n}\}_{n=1}^N$ are defined as
 \begin{equation} \label{MR_1}
 \mathcal{Y}_n= {\gamma _n}\left( \mathcal{X}_n - \bm{\mathcal{X}}_{n - 1}^ T \bm{\mathcal{C}}_{n - 1}^{ - 1}{\mathbf{c}}_n\right) /\sqrt{\mathcal{S}_n}, \mathcal{Y}_1={\gamma _1}\mathcal{X}_1 /\sqrt{\mathcal{C}_{1,1}},
 \end{equation}
  where ${\gamma _n}=\pm 1$,
  then the random variables $\{\mathcal{Y}_{n}\}_{n=1}^N$ can be produced by applying the Gram-Schmidt process to the random variables $\{\mathcal{X}_{n}\}_{n=1}^N$, and for any $1 \le n, m \le N$, the following equalities hold
 \begin{equation} \label{MainResult_1}  
 \begin {split}
 {\mathop{\mathrm {cov}}} ({\mathcal{Y}_n},{\mathcal{Y}_m}) &=0, \mbox{ if } n \ne m;\\
 {\mathop{\mathrm {cov}}} ({\mathcal{Y}_n},{\mathcal{Y}_n})  &=1,  \mbox{ if } n = m.
 \end {split}
 \end{equation} 
\end{theorem}
(Partially similar theorem can be found in \cite[Statement 1]{TrifonovKharinRus} or \cite[Statement 1]{TrifonovKharin1}.)

Next, suppose that random variables $\{{\mathcal{X}}_{j}\}_{j=1}^n$ have the following structure
	\begin{equation*}
		{\mathcal{X}}_{j}=\sum\limits_{t=1}^{N_s}x(t){s_j}(t,\bm{\theta}_j).
	\end{equation*}
Here
$x(t) = \sum\limits_{i = 1}^{{n_0 }}{a_{0i}} {{s_i}(t,{{\bm{\theta}}_{0i}})} + \xi(t)$,  ${s_i}(t,\bm{\theta}_{i})$ is a signal with a vector of parameters $\bm{\theta}_{i}$, $a_{0i} \in \mathbb{R}$, and $\xi(t)$ is the AWGN.
\begin{theorem} \label{zeroE}
If $n>n_0$ and $\bm{\theta}_{i}=\bm{\theta}_{0i}$ for all $i$ then
\begin{equation*}
\mathrm{E}\left(\mathcal{Y}_{0n}\right)=\mathrm{E}\!\left(\mathcal{X}_{0n} - \bm{\mathcal{X}}_{0n - 1}^ T \bm{\mathcal{C}}_{0n - 1}^{ - 1}{\mathbf{c}}_{0n}\right)=0,
\end{equation*}
where $\mathrm{E}$ denotes an expectation, ${\mathcal{X}}_{0j}=\sum\limits_{t=1}^{N_s}x(t){s_j}(t,\bm{\theta}_{0j})$, ${\mathcal{C}_{0i,j}} = {\mathop{\mathrm {cov}}} ({\mathcal{X}_{0i}},{\mathcal{X}_{0j}})$, ${\bm{\mathcal{C}}_{0n}} = (\mathcal{C}_{0i,j})_{i,j=1}^n$, $\bm{\mathcal{X}}_{0n}=\left( {\mathcal{X}}_{0j}\right)_{j=1}^n$. 
\end{theorem}
(This result is a generalization of \cite[Eq. 40]{TrifonovKharinRus})
\begin{IEEEproof}
Please see Appendix~\ref{Proof3}.
\end{IEEEproof}

\section{Proofs}
\subsection{Proof of the Theorem \ref{NewfGS}} \label{Proof1}
The case $n=1$ is obvious. Let us consider the case $n>1$.
Using the non-iterative formula for the Gram-Schmidt process from \cite{MTG}, we get
\begin{equation} \label{Th2_1}
\tilde B_n =  \frac{1}{\sqrt{\left|\mathbf{G}_{n-1}\right|\left|\mathbf{G}_{n}\right|}} \left| {\begin{array}{cccc}
{G_{1,1}}& \ldots &{G_{1,n - 1}}&{A_1}\\
 \vdots & \ddots & \vdots & \vdots \\
{G_{n - 1,1}}& \ldots &{G_{n - 1,n - 1}}&{A_{n - 1}}\\
{G_{n,1}}& \ldots &{G_{n,n - 1}}&{A_n}
\end{array}} \right|,
\end{equation}
where $\left|\cdot \right|$ denotes the determinant of a matrix. The determinant in the second factor at the right-hand side of \eqref{Th2_1} has to be formally expanded along the last column.

Next, using block matrices, we can represent \eqref{Th2_1} as
\begin{equation} \label{Th2_2}
\tilde B_n =\sqrt{\frac{\left|\mathbf{G}_{n-1}\right|}{\left|\mathbf{G}_{n}\right|}} \left( A_n + \frac{1}{\left|\mathbf{G}_{n-1}\right|}\sum\limits_{i=1}^{n-1}(-1)^{i+n}{A_i} {\begin{vmatrix} {\mathbf{G}_{n-1}^{\{i\} }}\\ {\mathbf{g}_n^H} \end{vmatrix}} \right),
\end{equation}
where $\mathbf{G}_{n-1}^{\{i\} }$ is the matrix obtained by deleting the $i$-th row from $\mathbf{G}_{n-1}$.

Let us expand the determinant of each block matrix  $ \begin{pmatrix} {\mathbf{G}_{n-1}^{\{i\} }}\\ {\mathbf{g}_n^H} \end{pmatrix} $ in \eqref{Th2_2} along the row $\mathbf{g}_n^H$
\begin{equation} \label{Th2_3}
\tilde B_n \!=\!\sqrt{\frac{\left|\mathbf{G}_{n-1}\right|}{\left|\mathbf{G}_{n}\right|}} \!\left(\!A_n\!+\!\sum\limits_{i=1}^{n-1}{\sum\limits_{k=1}^{n-1}\!{(-1)^{i+k}A_iM_{i,k}^{[n-1]}G_{k,n}}}\!\right)\!\!,
\end{equation}
where $M_{i,k}^{[n-1]} = - {\tilde M}_{i,k}^{[n-1]}/\left|\mathbf{G}_{n-1}\right|$ and ${\tilde M}_{i,k}^{[n-1]}$ is the $i,k$-th minor of $\mathbf{G}_{n-1}$, i.e.,  ${\tilde M}_{i,k}^{[n-1]}$ is the determinant of the submatrix that is obtained by deleting the $i$-th row and the $k$-th column from $\mathbf{G}_{n-1}$.

Now let us rewrite $B_n$ \eqref{MainFormula} using $M_{i,k}^{[n-1]}$:
\begin{equation} \label{Th2_4}
{B_n}\!=\!\left(\! A_n+\sum\limits_{i=1}^{n-1}{\sum\limits_{k=1}^{n-1}{(-1)^{i+k}A_iM_{i,k}^{[n-1]}G_{k,n}}}\!\right)\!/\!\sqrt {S_n}. 
\end{equation}

Comparing \eqref{Th2_3} and \eqref{Th2_4}, we get
\begin{equation} \label{Th2_5}
{\tilde B_n}\sqrt{\frac{\left|\mathbf{G}_{n}\right|}{\left|\mathbf{G}_{n-1}\right|}} = {B_n}{\sqrt {S_n}}.
 \end{equation}

Now to complete the proof of Theorem \ref{NewfGS} we need to prove the following equality:
\begin{equation} \label{Th2_6}
\left\langle {\tilde B_n},{\tilde B_n} \right\rangle = \left\langle {B_n},{B_n} \right\rangle. 
 \end{equation}
 It is well known that $\left\langle {\tilde B_n},{\tilde B_n} \right\rangle=1$ (see \cite{MTG}). Equality $\left\langle {B_n},{B_n} \right\rangle=1$ can be proved by direct calculations, thus we assume that equality \eqref{Th2_6} holds.
 
Comparing \eqref{Th2_5} and \eqref{Th2_6}, we get proof of Theorem \ref{NewfGS}.


\subsection{Proof of the Theorem \ref{zeroE}}\label{Proof3}
Firstly, the following result can be obtained by direct calculations: 
$\mathop{\mathrm{cov}}\left( {\mathcal{X}}_{i}, {\mathcal{X}}_{j}\right)=\sum_{t=1}^{N_s}{s_i}(t,\bm{\theta}_i){s_j}(t,\bm{\theta}_j). $

If $n > n_0$ and $\bm{\theta}_{i}=\bm{\theta}_{0i}$ for all $i$, then one can find that
\begin{equation} \label{ap3}
\begin{split}
\mathrm{E}\left(\bm{\mathcal{X}}_{0n-1}\right)&=\left(\sum\limits_{i=1}^{n_0}\sum\limits_{t=1}^{N_s}{a_{0i}}{s_i}(t,\bm{\theta}_{0i}){s_j}(t,\bm{\theta}_{0j})\right)_{j=1}^{n-1} \\
&=\left(\sum\limits_{i=1}^{n_0}a_{0i}\mathcal{C}_{0i,j}\right)_{j=1}^{n-1}=\left( \mathbf{a}_{0n_0}^T\bm{\mathcal{C}}_{0n - 1}\right)^T, 
\end{split}	
\end{equation}
where $\left( \mathbf{a}_{0n_0}\right)_i = \left\{ \!\! \begin{array}{l}
a_{0i}, \mbox{ if } i \le n_0,\\
0, \,\,\,\,\, \mbox{ if } i > n_0.
\end{array} \right.$

Equation \eqref{ap3} allows to prove Theorem \ref{zeroE} as follows
\begin{equation*}
\begin{split}
\mathrm{E}\left(\mathcal{Y}_{0n}\right)=&\mathrm{E}\left(\mathcal{X}_{0n}\right) - \mathrm{E}\left(\bm{\mathcal{X}}_{0n - 1}\right)^T \bm{\mathcal{C}}_{0n - 1}^{ - 1}{\mathbf{c}}_{0n}\\
=&\left(\mathbf{a}_{0n_0}^T{\mathbf{c}}_{0n} - \mathbf{a}_{0n_0}^T\bm{\mathcal{C}}_{0n - 1}\bm{\mathcal{C}}_{0n - 1}^{ - 1}{\mathbf{c}}_{0n} \right) =0.
\end{split}
\end{equation*}

\subsection{Proof of the Theorem \ref{unknownNoiseOrh}} \label{proofUNO}
According to Theorem \ref{randOrh} and the representation of the Gram-Schmidt process \eqref{MainFormula} (see Theorem \ref{NewfGS}), the random variables $\{l_{\sigma i}^{*2}\}$ are obtained sequentially by projecting the data onto the subspace orthogonal to the frequencies $\bm{\omega}_1^*, \dots, \bm{\omega}_{i-1}^*$. Let $\bm{\mathcal{E}}_{i-1}$ denote the residual vector in this orthogonal subspace at the $i$-th step. Using \eqref{LFrSigDE}, one can find that the random variable $l_{\sigma i}^{*2}$ depends on the direction of $\bm{\mathcal{E}}_{i-1}$ (defined by $\bm{\mathcal{E}}_{i-1}/\|\bm{\mathcal{E}}_{i-1}\|$), and does not depend on $\|\bm{\mathcal{E}}_{i-1}\|^2$.
Next, using \eqref{MainFormula}, we obtain $\|\bm{\mathcal{E}}_{i-1}\|^2 = \sum_{t=1}^{N_s} x^2(t) - \sum_{k=1}^{2i-2} \sigma_0^2 \mathcal{Y}_k^2$. The random variable $l_{\sigma i}^{*2}$ can be represented as
\begin{equation} \label{btatran}
l_{\sigma i}^{*2} = \frac{\mathcal{Y}_{2i-1}^2 + \mathcal{Y}_{2i}^2}{\frac{1}{\sigma_0^2}\sum_{t=1}^{N_s} x^2(t) - \sum_{k=1}^{2i-2} \mathcal{Y}_k^2} = \frac{\mathcal{V}_i}{\mathcal{V}_i + \mathcal{Q}_i},
\end{equation}
where $\mathcal{V}_i = \mathcal{Y}_{2i-1}^2 + \mathcal{Y}_{2i}^2$ and $\mathcal{Q}_i = \frac{1}{\sigma_0^2}\sum_{t=1}^{N_s} x^2(t) - \sum_{k=1}^{2i} \mathcal{Y}_k^2$. Using Cochran's theorem, one can find that the random variables $\mathcal{V}_i$ and $\mathcal{Q}_i$ are independent.

For $i > \nu_0$, the expected value of the residual vector $\bm{\mathcal{E}}_{i-1}$ is zero ($\mathrm{E}(\bm{\mathcal{E}}_{i-1}) = \mathbf{0}$). Thus, $\mathcal{V}_i$ follows a central chi-square distribution with $2$ degrees of freedom ($\mathcal{V}_i \sim \chi^2(2)$), and $\mathcal{Q}_i$ follows a central chi-square distribution with $N_s - 2i$ degrees of freedom ($\mathcal{Q}_i \sim \chi^2(N_s - 2i)$). Using these results and representation \eqref{btatran}, one can conclude that the random variables $\{l_{\sigma i}^{*2}\}_{i>\nu_0}$ follow the Beta distribution with parameters $\alpha = 1$ and $\beta = (N_s - 2i)/2$ (see the CDF in \eqref{Beta_CDF}). 

Furthermore, if $i>\nu_0$, then $\bm{\mathcal{E}}_{i-1}$ is a zero-mean isotropic Gaussian vector, and thus its direction is independent of its squared norm $\|\bm{\mathcal{E}}_{i-1}\|^2$ \cite{Lehmann2005}. Consequently, $l_{\sigma i}^{*2}$ is independent of $\|\bm{\mathcal{E}}_{i-1}\|^2$. On the one hand, random variables $l_{\sigma 1}^{*2}, \ldots, l_{\sigma, i-1}^{*2}$ influence the vector $\bm{\mathcal{E}}_{i-1}$ only through the norm $\|\bm{\mathcal{E}}_{i-1}\|$, on the other hand, the random variable $l_{\sigma i}^{*2}$ is independent of this norm. This proves properties 1, 2, and 3.

Let us prove property 5. The CDF of $\Delta \tilde{L}_{\sigma i}^* = -\frac{N_s}{2}\ln(1 - l_{\sigma i}^{*2})$ from \eqref{Main_Rep_Ksigma} for any $v \ge 0$ can be represented as
\begin{equation*}
\begin{split}
F_{\Delta L, i}^*(v) &= \Pr\left(-\frac{N_s}{2}\ln(1 - l_{\sigma i}^{*2}) < v\right) \\
&= \Pr\left(l_{\sigma i}^{*2} < 1 - e^{-2v/N_s}\right).
\end{split}
\end{equation*}
Substituting this result into the Beta CDF \eqref{Beta_CDF}, we obtain
\begin{equation*}
\begin{split}
F_{\Delta L, i}^*(v) &= 1 - \left(1 - \left(1 - e^{-2v/N_s}\right)\right)^{\frac{N_s - 2i}{2}}\\
& = 1 - \exp\left(-v \frac{N_s - 2i}{N_s}\right),
\end{split}
\end{equation*}
which yields the exponential distribution in \eqref{Exp_CDF}.

Finally, let us prove property 4. If $i = \nu_0$, then the expected value of the residual vector $\bm{\mathcal{E}}_{\nu_0-1}$ is non-zero, and $\mathcal{V}_{\nu_0}$ is a non-central chi-square random variable ($\mathcal{V}_{\nu_0} \sim \chi^2(2, \lambda_{\nu_0})$). The random variable $\mathcal{Q}_{\nu_0}$ depends only on the noise, and thus follows a central chi-square distribution ($\mathcal{Q}_{\nu_0} \sim \chi^2(N_s - 2\nu_0)$). Taking into account these results, one can conclude that the ratio \eqref{btatran} of these independent variables defines the singly non-central Beta distribution.





\bibliographystyle{IEEEtran}
\bibliography{Kharin2020bib}
%
\begin{IEEEbiography}[{\includegraphics[width=1in,height=1.25in,clip,keepaspectratio]{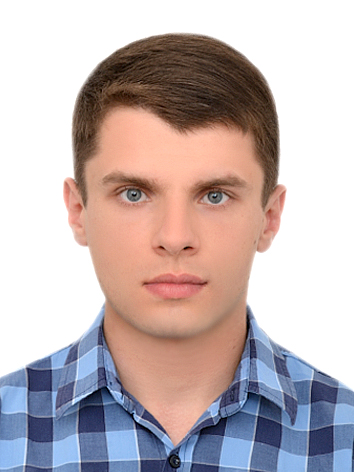}}]{Aleksandr Kharin}
was born in Voronezh, Russia in 1989. He received the B.S., M.S., and Ph.D. degrees from the Department of Radiophysics, Voronezh State University, Voronezh, Russia, in 2010, 2012, and 2016, respectively, all in radiophysics. His main interests are statistical signal processing, estimation theory, applied probability, and applied statistics. 

Dr. Kharin is IEEE Member since 2016.
\end{IEEEbiography}

\end{document}